\documentclass{sig-alternate}

\usepackage{amsmath}
\usepackage{amsfonts}
\usepackage{amssymb}
\usepackage{balance}
\usepackage{enumitem}

\usepackage{floatrow}
\usepackage{graphicx}
\usepackage[T1]{fontenc}
\usepackage[font=small,labelfont=bf,tableposition=top]{caption}
\DeclareCaptionLabelFormat{andtable}{#1~#2  \&  \tablename~\thetable}

\usepackage{caption}
\newfloatcommand{capbtabbox}{table}[][\FBwidth]

\begin{document}

\CopyrightYear{2017}
\setcopyright{acmlicensed}
\conferenceinfo{WSDM 2017,}{February 06 - 10, 2017, Cambridge, United Kingdom}
\isbn{978-1-4503-4675-7/17/02}\acmPrice{\$15.00}
\doi{http://dx.doi.org/10.1145/3018661.3018733}

\title{Evolution of Ego-networks\\in Social Media with Link Recommendations}

\numberofauthors{2}
\author{
\alignauthor
Luca Maria Aiello\\
       \affaddr{Nokia Bell Labs}\\
       \affaddr{Cambridge, United Kingdom}\\
       \email{luca.aiello@nokia-bell-labs.com}
\alignauthor
Nicola Barbieri\\
       \affaddr{Tumblr}\\
       \affaddr{New York, NY, USA}\\
       \email{barbieri@yahoo-inc.com}
}

\date{12 Dec 2016}

\maketitle
\begin{abstract}
Ego-networks are fundamental structures in social graphs, yet the process of their evolution is still widely unexplored. In an online context, a key question is how link recommender systems may skew the growth of these networks, possibly restraining diversity. To shed light on this matter, we analyze the complete temporal evolution of 170M ego-networks extracted from Flickr and Tumblr, comparing links that are created spontaneously with those that have been algorithmically recommended. We find that the evolution of ego-networks is bursty, community-driven, and characterized by subsequent phases of explosive diameter increase, slight shrinking, and stabilization. Recommendations favor popular and well-connected nodes, limiting the diameter expansion. With a matching experiment aimed at detecting causal relationships from observational data, we find that the bias introduced by the recommendations fosters global diversity in the process of neighbor selection. Last, with two link prediction experiments, we show how insights from our analysis can be used to improve the effectiveness of social recommender systems.
\end{abstract}

%
%



\section{Introduction} \label{sec:intro}

Ego-centric social networks (ego-networks) map the interactions that occur between the social contacts of individual people. Because they provide the view of the social world from a personal perspective, these structures are fundamental information blocks to understand how individual behaviour is linked to group life and societal dynamics. Despite the growing availability of interaction data from online social media, little research has been conducted to unveil the structure and evolutionary dynamics of ego-networks~\cite{arnaboldi12analysis,kikas13bursty}. In an online context, people expand their social circles also as a result of automatic recommendations that are offered to them, which makes it harder to disentangle spontaneous user behavior from algorithmically-induced actions.

We aim to provide an all-round description about how ego-networks are formed and how automated contact recommendations might bias their growth. We do so by analyzing the full longitudinal traces of 170M ego-networks from Flickr and Tumblr ($\S$\ref{sec:dataset}), answering several open research questions about the shape of their boundaries, their community structure, and the process of neighbor selection in time ($\S$\ref{sec:analysis}). The richness of the data we study allows for the identification of those Tumblr links that have been created as a result of recommendations served by the platform, which positions us in an unique standpoint to investigate the impact of link recommender systems on the process of network growth.

Some of our key findings are:
\vspace{-1pt}
\begin{itemize}[leftmargin=*]
\itemsep1pt 
	\item The backbone of a typical ego-network is shaped within the initial month of a node's activity and within the first $\sim$50-100 links created. In that period, new contacts are added in larger batches and the main communities emerge. Unlike global social networks, whose diameter shrinks in time, the average distance between nodes in ego-networks expands rapidly and then stabilizes.
	\item The selection criteria of new neighbors change as new contacts are added, with popular contacts being more frequently followed in earlier stages of the ego's life, and friends-of-friends being selected in later stages. The neighbor selection is also heavily driven by the ego-network's community structure, as people tend to grow different sub-groups sequentially, with an in-depth exploration strategy.
	\item The link recommender system skews the process of ego-network construction towards more popular contacts but at the same time restraining the growth of its diameter, compared to spontaneous behavior. With a matching experiment aimed at detecting causal relationships from observational data, we find that the bias introduced by the recommendations fosters diversity: people exposed to recommendations end up creating pools of contacts that are more different from each other compared to those who were not exposed.
\end{itemize}

The outcomes of our analysis have theoretical implications in network science and find direct application in link recommendation and prediction tasks. We run a prediction experiment ($\S$\ref{sec:prediction}) to show that simple temporal signals could be crucial features to improve link prediction performance, as the criteria of ego-network expansion vary as the ego grows older. In a second experiment, we test the algorithmic capability to tell apart spontaneous links from recommendation-induced ones. This ability opens up the way to train link predictors that mitigate existing algorithmic biases by suggesting links whose properties better adhere to the natural criteria that people follow when connecting to others.

\section{Related work} \label{sec:related}

\noindent \textbf{Structure and dynamics of social networks.}
For decades, network science research has explored extensively the structural and evolutionary properties of online social graphs and of the communities they encompass~\cite{garton97studying,barabasi02evolution,kossinets06empirical,backstrom06group,palla07quantifying,mislove07measurement,wilson09user,yang11patterns,tan15all}, unveiling universal patterns of their dynamics. Individual connectivity and activity are broadly distributed~\cite{mislove07measurement,ugander11anatomy}; the creation of new links is driven by reciprocation, preferential attachment~\cite{mislove08growth}, triangle closure~\cite{leskovec08microscopic}, and homophily~\cite{aiello12tweb,yuan14exploiting}. Globally, the number of edges in a social network grows superlinearly with its number of nodes, and the average path length shrinks with the addition of new nodes~\cite{leskovec05graphs}, after an initial expansion phase~\cite{ahn07analysis}. The regular patterns that drive the link creation process have enabled the development of accurate methods for link prediction and recommendation~\cite{hasan11survey} based on either local~\cite{libennowell03link} or global structural information~\cite{bahmani10fast,backstrom11supervised,shin15tumblr}. Fine-grained temporal traces of user activity in online social platforms opened up new avenues to investigate in detail the impact of time on network growth~\cite{zignani14link}. For example, the relationship between the node age and its connectivity has been measured in several online social graphs including Flickr~\cite{leskovec08microscopic,yin11link}.

\vspace{4pt}
\noindent \textbf{Ego-networks.}  
To date, not much research has been conducted on how nodes build their \textit{local} social neighborhoods in time. Research done by Aranboldi et al. has looked into ego-networks of online-mediated relationships including the Facebook friendship network~\cite{arnaboldi12analysis} and the Twitter follow graph~\cite{arnaboldi13ego,arnaboldi16ego}, as well as professional relationships such as Google Scholar's co-authorship network~\cite{arnaboldi16analysis}. Using community detection, hierarchical clusters are discovered, in agreement with Robin Dunbar's theory on the hierarchical arrangement of social ego-circles~\cite{zhou05discrete}. Similar findings have been confirmed by independent studies on the Facebook network~\cite{desalve16impact}. In the attempt of comparing the  properties of the global network with those of ego-networks, recent studies found that local structural attributes are characterized by local biases~\cite{gupta15structural} that are direct implications of the friendship paradox~\cite{feld91friends}. Multiple techniques have been proposed to discover social or topical sub-groups within ego-networks~\cite{weng14topic,mcauley14discovering,muhammad15duke,biswas15community}, but with little attention to the dynamics of their growth. Kikas et al. conducted one of the few studies touching upon the the temporal evolution of ego-networks, using a dataset of Skype contacts~\cite{kikas13bursty}. They find that most edges are added in short bursts separated by long inactivity intervals. 

\vspace{4pt}
\noindent \textbf{Effect of social recommender systems.}
In the past years, computer scientists developed increasingly effective contact recommender systems for online social media~\cite{gupta13wtf}. Only recently, the community has adopted a more critical standpoint with respect to the \textit{effects} that those recommendations may have on the collective user dynamics. Algorithms based on network proximity are better suited to find contacts that are already known by the user, whereas algorithms based on similarity of user-generated content are stronger at discovering new friends~\cite{chen09make}. Surveys administered to members of corporate social networks revealed that contact recommendations with high number of common neighbors are usually well-received~\cite{daly10network}. Recommendation-induced link creations have a substantial effect on the growth of the social graph; for example, the introduction of the ``people you may know'' service in Facebook increased considerably the number of links created and the ratio of triangle closures~\cite{zignani14link}. A recent study on Twitter compared the link creation activity before and after the introduction of the ``Who To Follow'' service~\cite{su16effect}, showing that popular nodes are those who most benefit from recommendations. On a wider perspective, the debate around recommender systems fostering or limiting access to novel information is still open. On one hand, recommenders may originate a filter bubble effect by providing information that increasingly reinforces existing viewpoints~\cite{pariser12filter}. On the other hand, recent research has pointed out that individual choices, more than the effect of algorithms, limit exposure to cross-cutting content~\cite{bakshy15exposure}. It has been argued that recommender systems have a limited effect in influencing people's free will. Observational studies on Amazon found that $75\%$ of click-throughs on recommended products would likely have occurred also in the absence of recommendations~\cite{sharma15estimating}. In the context of link recommendation systems, a key open question is how they affect ego-network diversity.

\section{Dataset and preliminaries}  \label{sec:dataset}

We study two social media platforms that differ in both scope and usage. The data includes only interactions between users who voluntarily opted-in for research studies. All the analysis has been performed in aggregate and on anonymized data.

\subsection{Tumblr} \label{sec:dataset:tumblr}

\begin{figure}[t!]
\begin{center}
\includegraphics[width=0.50\columnwidth]{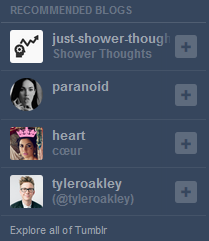}
\caption{Tumblr's contact recommender system.}
\label{fig:tumblr_recs_screenshot}
\end{center}
\end{figure}

Tumblr is a popular social blogging platform. The types of user-generated content range from simple textual messages to multimedia advertising campaigns~\cite{chang14what,grbovic15gender}. Users might own multiple blogs, but for the purpose of this study we consider blogs as users, and we will use the two terms interchangeably. Users receive updates from the blogs they follow; the following relationship is directional and might not be reciprocated. In Tumblr, 326 million blogs and 143 billion posts have been created\footnote{\url{https://www.tumblr.com/about} (Dec 2016)} since the release of the platform to the public in July 2007. We extracted a large random sample of the social network in October 2015, which includes almost $7B$ follow links created between $130M$ public blogs over approximately 8 years. All the social links are marked with the exact timestamps of their creation, which allows for a fine-grained longitudinal analysis of the network evolution.

In October 2012, Tumblr launched a new version of its \textit{recommended blogs} feature. On the web interface, a shortlist of four recommended blogs is displayed in a panel next to the user's feed (Figure~\ref{fig:tumblr_recs_screenshot}). Users can get more recommendations by clicking on ``explore''. At every page refresh, the shortlist may change according to a randomized reshuffling strategy that surfaces new recommended contacts from the larger pool. Tumblr's link recommendation algorithm is not publicly disclosed, but it considers a mixture of two signals: topical preferences and network structure. The user's tastes are estimated since the onboarding phase, in which registrants are asked to indicate their preference on a set of pre-determined topics organized in a taxonomy (e.g., sports, football). The topical profile helps to overcome the cold-start problem. As the number of contacts grows, new blogs are recommended following the triangle closure (friend-of-a-friend) principle. 

For all the links created after January 2015, we can reliably estimate if they have been created as an effect of a recommendation. This information is inferred by combining the log of recommendation impressions (i.e., when recommendations are visualized by the user) with the log of link creations. When the link is created shortly after the recommendation is displayed, we count the link creation as triggered by a recommendation.

\begin{figure}[t!]
\begin{center}
\includegraphics[width=0.49\columnwidth]{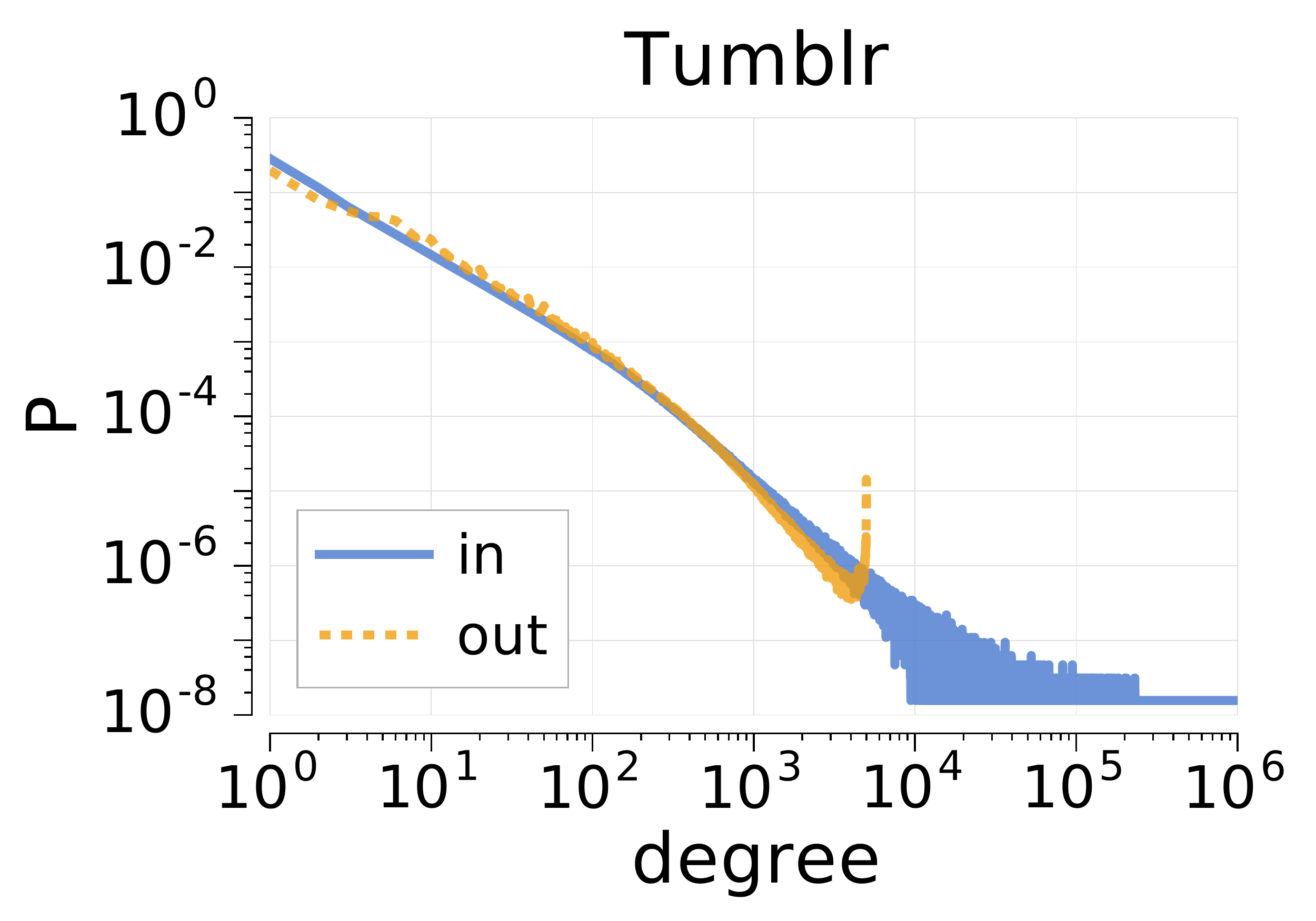}
\includegraphics[width=0.49\columnwidth]{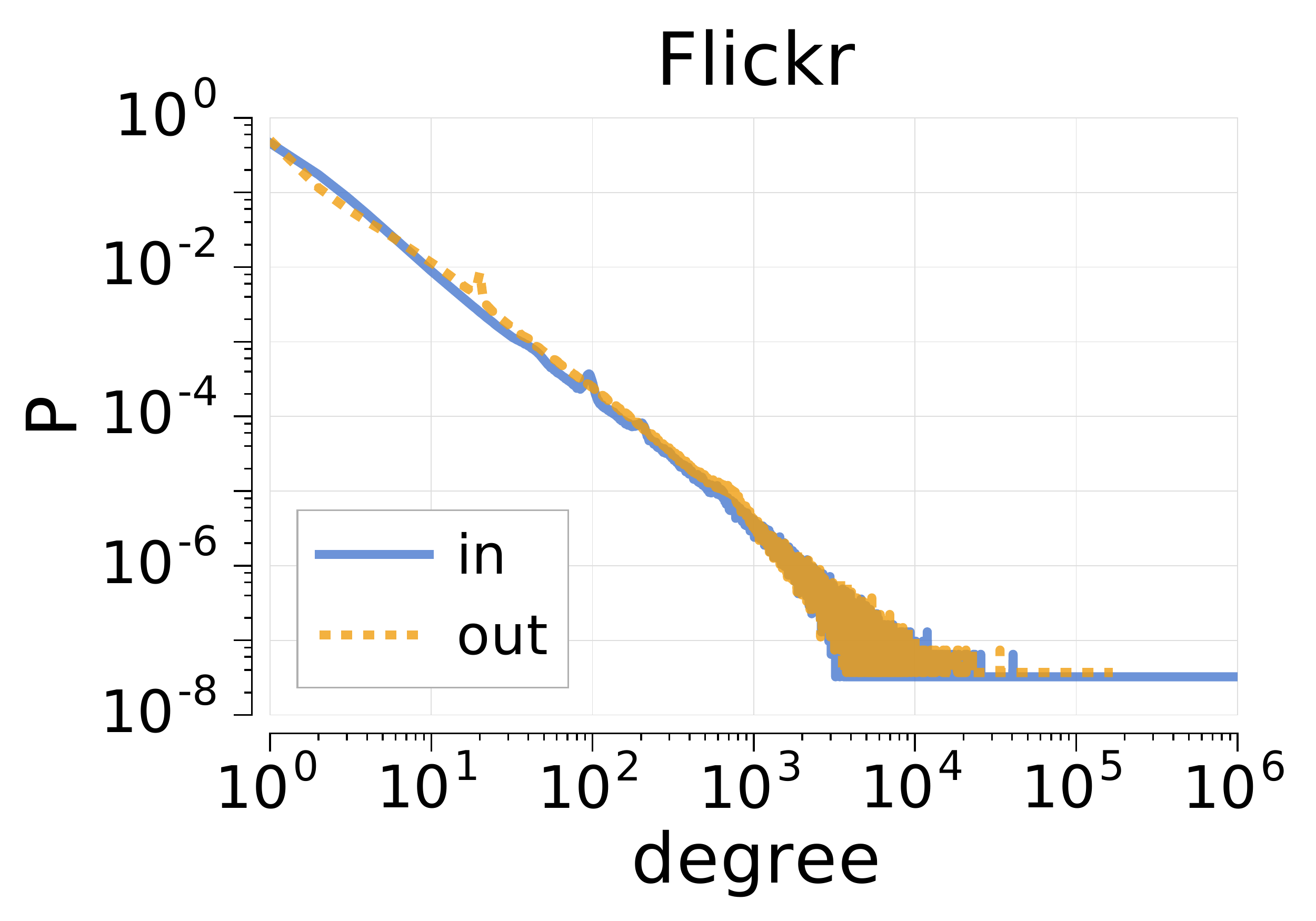}
\caption{Degree distributions in Tumblr ($\mu_{in}=108$, $\mu_{out}=58$) and Flickr ($\mu_{in}=19$, $\mu_{out}=21$).}
\label{fig:degree_distr}
\vspace{2mm}
\end{center}
\end{figure}

\subsection{Flickr} \label{sec:dataset:flickr}

Flickr is a popular photo-sharing platform in which users can upload a large amount (up to 1 TB) of pictures and share them with friends. Users can establish directed social links by following other users and get updates on their activity. Since its release in February 2004, the platform has gathered almost 90 million registered members who upload more than 3.5 million new images daily\footnote{\url{http://www.theverge.com/2013/3/20/4121574/flickr-chief-markus-spiering-talks-photos-and-marissa-mayer}}. We collected a sample of the follower network composed by the nearly $40M$ public Flickr profiles that are opted-in for research studies and by the $500M+$ links that connect them. Links carry the timestamp of their creation and they span approximately 12 years ending March 2016. Similar to Tumblr, Flickr has a contact recommendation module. However, we do not have access to recommendation data and we cannot measure their effect on the link creation process.

\begin{figure}[t!]
\begin{center}
\includegraphics[width=0.95\columnwidth]{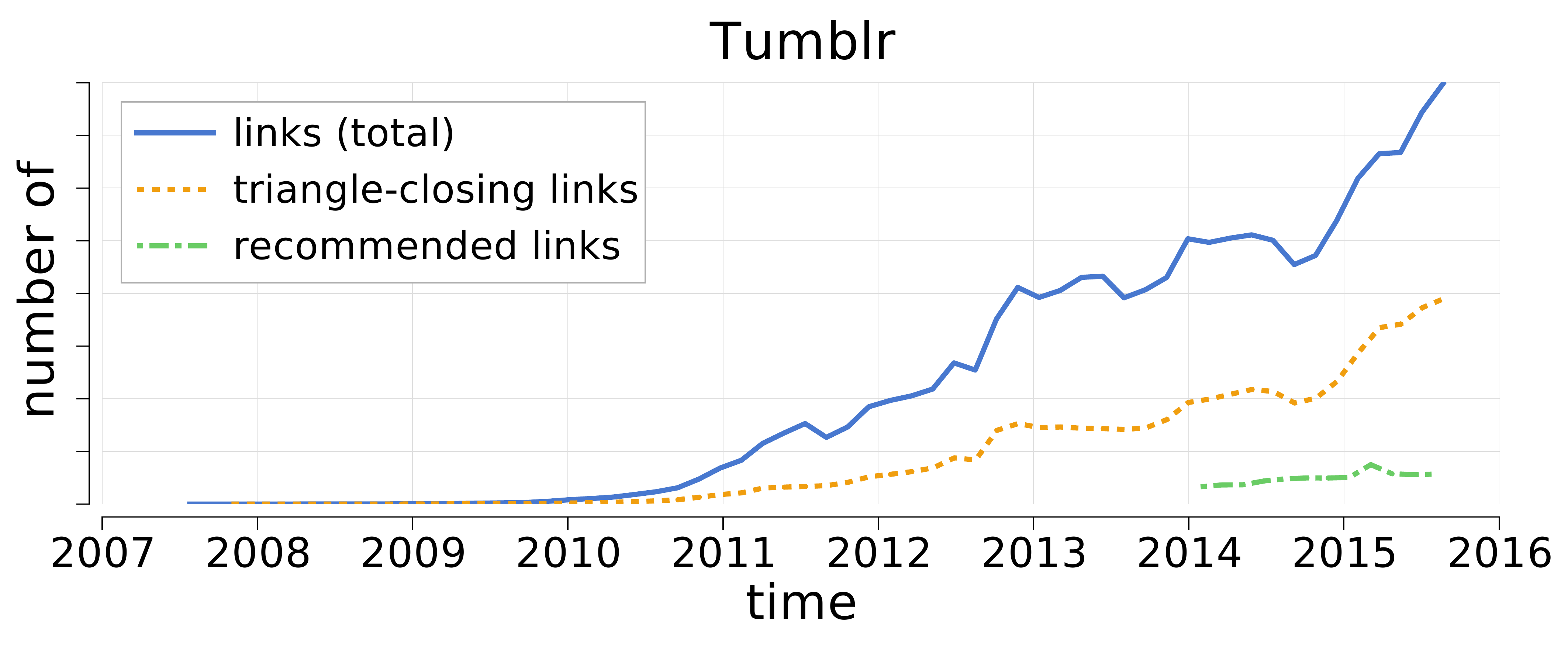}\\
\includegraphics[width=0.95\columnwidth]{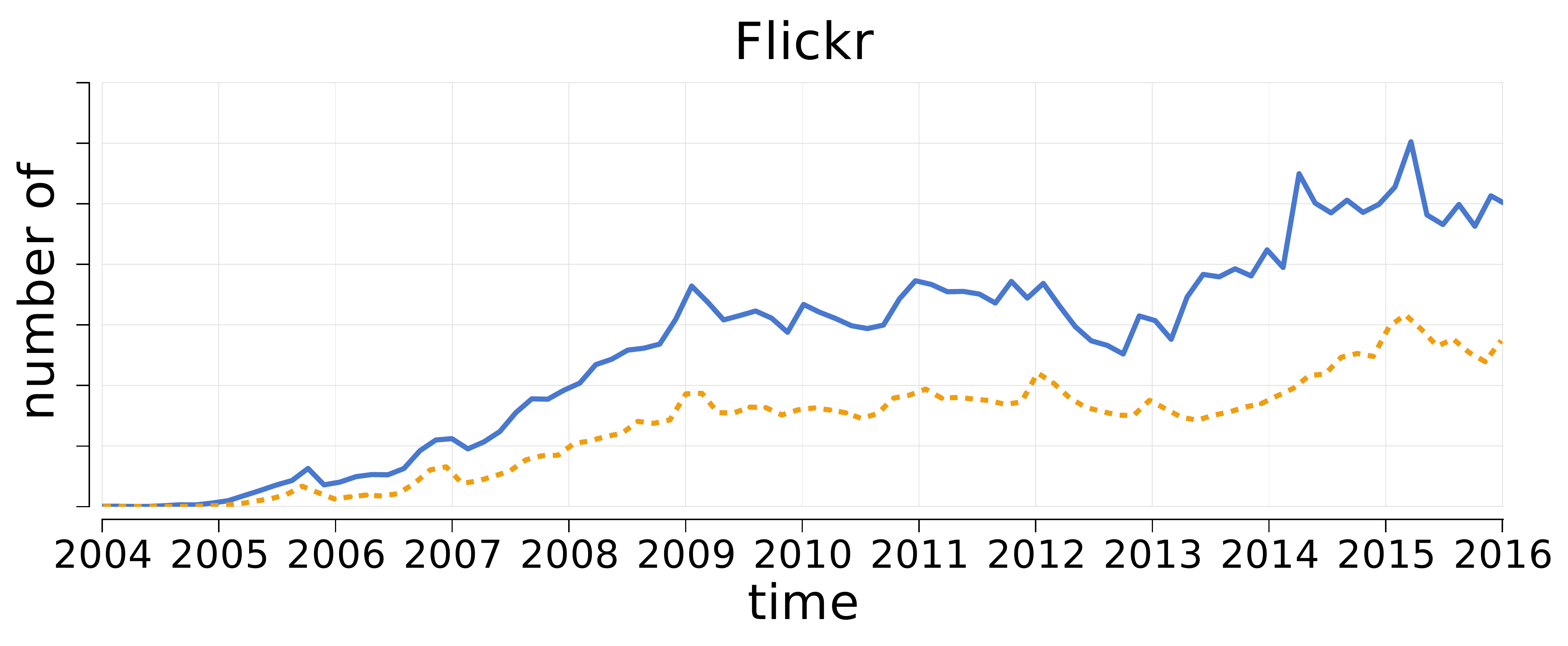}
\caption{Number of links, triangle-closing links, and recommended links (Tumblr only) created each day, by the nodes in our sample, during the whole lifespan of the platforms.}
\label{fig:timelines}
\end{center}
\end{figure}

\subsection{Concepts and notation} \label{sec:dataset:notation}

\noindent \textbf{Graph and ego-network.} Consider a follower graph $\mathcal{G}$ composed by a set of nodes $\mathcal{N}$ and a set of directed edges\footnote{We will use the terms \textit{directed edges}, \textit{edges}, and \textit{links} interchangeably to indicate a directional connection between nodes.} $\mathcal{E} \in \mathcal{N} \times \mathcal{N}$. When building the follower graph, we draw an edge from node $i$ to node $j$ if $i$ has followed $j$ at any time. The \textit{ego-network} $\mathcal{G}_i$ of node $i \in \mathcal{N}$ is the subgraph induced by $i$'s out-neighbors $\Gamma_{out}(i)$~\cite{freeman82centered}. Formally: $\mathcal{G}_i = (\mathcal{N}_i, \mathcal{E}_i)$, where $\mathcal{N}_i = \Gamma_{out}(i)$, and $\mathcal{E}_i = \{ (j,l) \in \mathcal{E} | $ $ j \in \Gamma_{out}(i) \wedge l \in \Gamma_{out}(i) \}$. Note that the ego-network does \textit{not} include the links between the ego $i$ and its neighbors.

\vspace{4pt}
\noindent \textbf{Structural graph metrics in time.} The temporal trace of link creations allows us to build a time graph~\cite{kumar03bursty} and to recover the structural properties of nodes and links at any point in time. The superscript $t$ applied to any indicator means that the metric refers to a snapshot of the graph at time $t$. For example, the neighbor set of node $i$ at time $t$ is denoted as $\Gamma^t(i)$ and its degree as $k^t(i)$. When studying the evolution of ego-networks in isolation, we will consider time on a discrete scale where each event corresponds to the $n^{th}$ node being added to the ego-network. We will use the letter $n$ to denote time passing on this discrete scale. All the graphs we consider are directed, so we use the definition of triangle closure adapted to directed graphs~\cite{romero10directed}: a new link created between $i$ and $j$ at time $t$ closes a directed triangle if $\exists l \in \mathcal{N} | l \in \Gamma^t_{out}(i) \wedge l \in \Gamma^t_{in}(j)$.

\vspace{4pt}
\noindent \textbf{Spontaneous vs. recommended links.} We distinguish links that are created for effect of a recommendation from those that are not. We call the links in the first group \textit{recommended} and the ones in latter \textit{spontaneous}.

\subsection{Data overview} \label{sec:dataset:overview}

The (in/out)degree distributions together with their average values ($\mu$) are shown in Figure~\ref{fig:degree_distr}. As expected, all distributions are broad, with values spanning several orders of magnitude. The out-degree distribution in Tumblr is capped at 5000 because the platform imposes an upper bound on the number of blogs a user can follow. On average, Tumblr users are more connected than Flickr users, with average in- and out-degree 5 and 3 times larger than Flickr, respectively.

Figure~\ref{fig:timelines} plots the number of links created over the course of the platforms' life. Both networks have experienced a noticeable growth. For Tumblr, we can plot the time series of recommended link creations occurred after January 2015. We also calculate the set of links that close at least one triangle. In Tumblr, the first sharp increase in the number of triangle-closing links is found between 2012 and 2013. That is determined by the introduction of a new link recommender system. A similar pattern has been observed in Facebook after the introduction of the ``people you may know'' module~\cite{zignani14link}. About $27\%$ of recommended links do not close any triangle: those are recommendations based on the user's topical profile only.

\section{Evolution of ego-networks} \label{sec:analysis}

\subsection{Diameter and connected components} \label{sec:analysis:diameter}

The growth of social networks is associated with three changes in their macroscopic structure: densification, diameter shrinking, and inclusion of almost all nodes in a single giant connected component~\cite{leskovec05graphs}. It is unknown whether the same properties hold at ego-network level.

\begin{figure}[t!]
\begin{center}
\includegraphics[width=0.49\columnwidth]{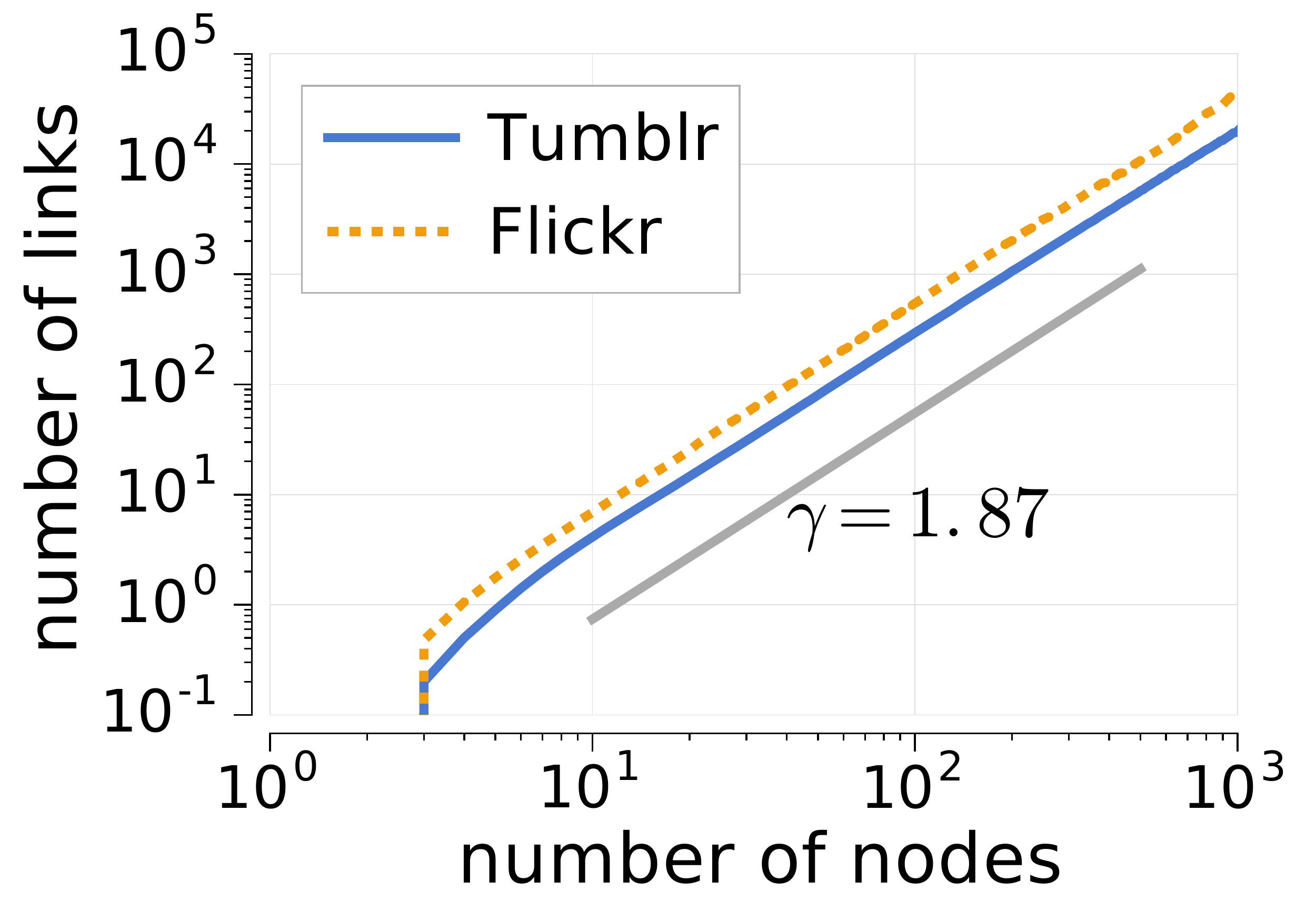}
\includegraphics[width=0.49\columnwidth]{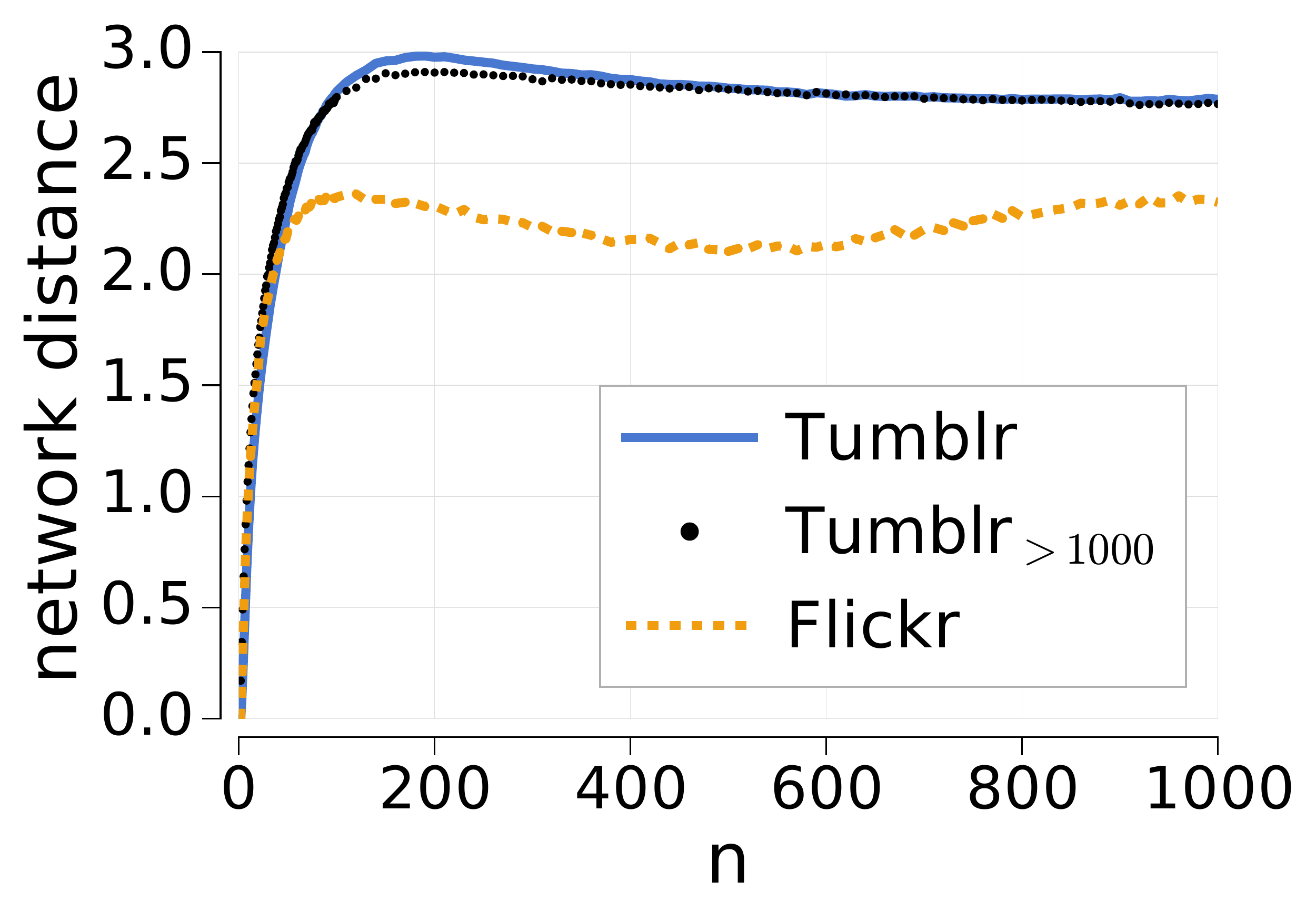}
\caption{Left: average number of ego-network links vs. number of nodes; best fitting power-law exponent is reported as reference. Right: average network distance after the $n^{th}$ node is added; the black dotted line is obtained considering only Tumblr ego-networks with at least 1000 nodes.}
\label{fig:density_diameter}
\end{center}
\end{figure}

\vspace{4pt}
\noindent \textbf{Q1: How do density, diameter, and component structure evolve as the ego-network grows?}

Like global networks, ego-networks become denser in time. Ego-networks obey a densification power law, for which the number of links scales superlinearly with the number of nodes $|\mathcal{E}_i| \sim |\mathcal{N}_i|^{\gamma}$ (Figure~\ref{fig:density_diameter}, left). The exponent that best defines the scaling in both platforms is $\gamma=1.87$.

\begin{figure}[t!]
\begin{center}
\includegraphics[width=0.49\columnwidth]{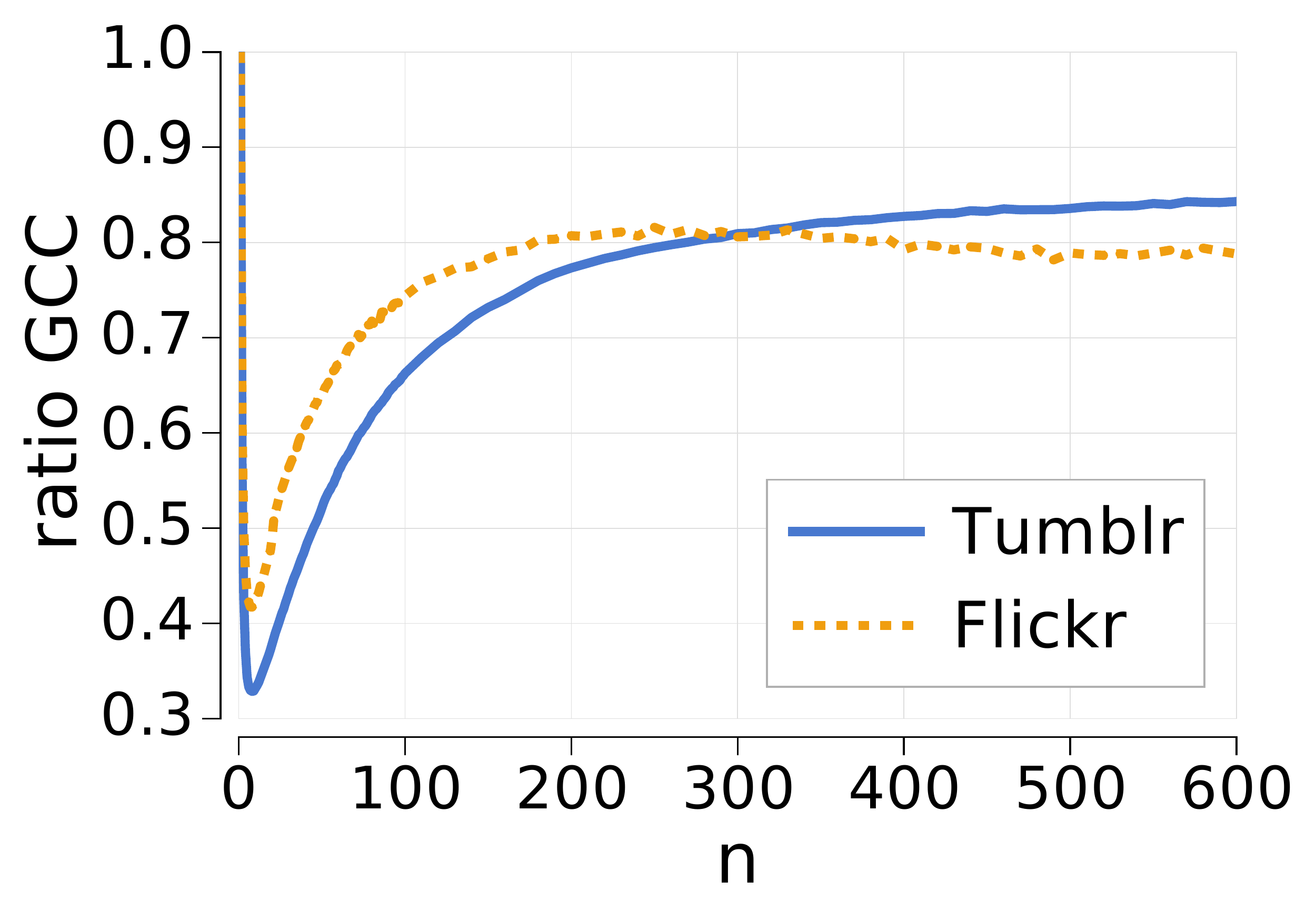}
\includegraphics[width=0.49\columnwidth]{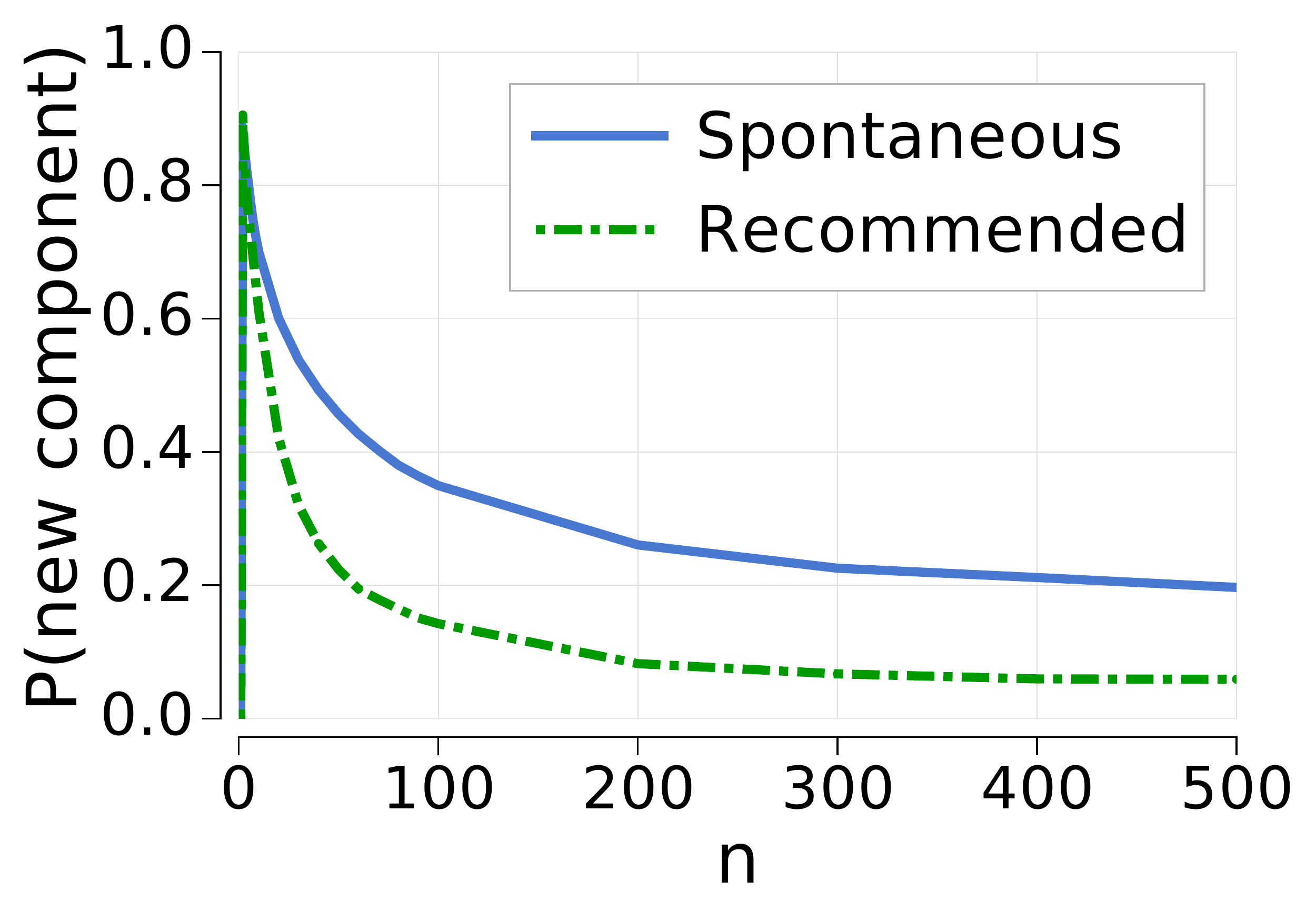}
\caption{Left: average ratio of nodes in the giant weakly connected component (GCC) as new nodes are added to the ego network. Right: probability that the $n^{th}$ node included in the ego-network spawns a new disconnected component, computed for spontaneous and recommended nodes.}
\label{fig:components}
\end{center}
\end{figure}

More surprisingly, densification does not always lead to the emergence of a single giant connected component covering the whole graph. On average\footnote{\small Results are qualitatively similar when considering the median.}, the largest component's size relative to the network size grows as new nodes join (Figure~\ref{fig:components}, left), but it stabilizes around $0.8$ for networks of 200 nodes or more. The number of components grows sublinearly with the number of nodes (not shown). More notably, the diameter shows little signs of shrinking. The \textit{network distance}, computed as the average distance between all pairs of nodes,\footnote{\small Computed on an undirected version on the graph. Similar results are obtained using diameter or effective diameter.} experiences a three-phases evolution (Figure~\ref{fig:density_diameter}, right). First, at the beginning of the ego-network life, it expands rapidly (\textit{exploration}); then, it starts shrinking slightly (\textit{consolidation}) before asintotically converging to a stable value (\textit{stabilization}). This trend is very different from the sharp diameter decline that characterizes social graphs. We speculate that the consolidation phase might be connected with the intrinsic human limitation to maintain large social groups, as theorized by Robin Dunbar~\cite{dunbar98grooming}. When the ego's social neighborhood exceeds the size that is cognitively manageable by a person (roughly, 150 to 200 individuals), a compensation effect might be triggered: new contacts are not anymore sought further away from the social circles that have been already established, putting an end to the exploration phase. This happens at $n=140$ in Flickr and $n=190$ in Tumblr, values that are compatible with Dunbar's theory.

The addition of recommended nodes has a different effect on the ego-network expansion, compared to spontaneous ones. New recommended contacts tend to be closer to existing ego-network members. At fixed network size, the addition of new recommended nodes increases the network distance $5\%$ less, on average, than a spontaneous node addition. Ego-networks with at least a recommended node have smaller network distance than ego-networks that grew fully spontaneously; this difference varies with the size, being only $2\%$ smaller for networks under 50 nodes up to $10\%$ smaller for networks with 200 nodes or more. Also, spontaneous nodes have far higher chances, compared to recommended ones, to spawn a new component disconnected from the rest of the ego-network (Figure~\ref{fig:components}, right).

\vspace{4pt}\noindent\textbf{Accounting for amalgamation effects.} To explore evolutionary trends of ego-networks, we rely on aggregate analysis: an indicator is measured on an ego-network when its $n^{th}$ node is added and then it is averaged across all ego-networks. Trends are discovered as $n$ grows (e.g., diameter in Figure~\ref{fig:density_diameter}, right). This approach may yield misleading results because averages computed at different values of $n$ are obtained from different sample sets. This problem is known as the \textit{Simpson's Paradox}~\cite{simpson51interpretation} and it is usually addressed by fixing the sample set~\cite{barbosa16averaging}. To account for it, every time we perform an evolutionary analysis as $n$ varies in $[1,n_{max}]$, we compare results obtained in two settings: the first using the full dataset and the latter considering only the subset of ego-networks that reached at least size $n_{max}$. The results are only slightly different across the two settings, for all the indicators analyzed. For the sake of brevity, we report just one example of such comparison. In Figure~\ref{fig:density_diameter} right, the diameter evolution for Tumblr ego-networks that reached at least size 1000 is very similar to the trend found when all ego-networks are considered.

\subsection{Popularity vs. similarity} \label{sec:analysis:predictors}

The process of link creation in online social networks is driven by two main factors: popularity (that leads to preferential attachment~\cite{mislove08growth}) and similarity (that leads to homophily~\cite{aiello12tweb}). At network scale, their relative weight in predicting the creation of new links might vary depending on the type of social network~\cite{hasan11survey}. At microscopic scale, it is still unclear how popularity and similarity impact the selection of new nodes in ego-networks, and how their relative importance varies in time.

\vspace{4pt}
\noindent \textbf{Q2: How do the criteria of neighbor selection change as the ego-network grows?}

\begin{figure}[t!]
\begin{center}
\includegraphics[width=0.49\columnwidth]{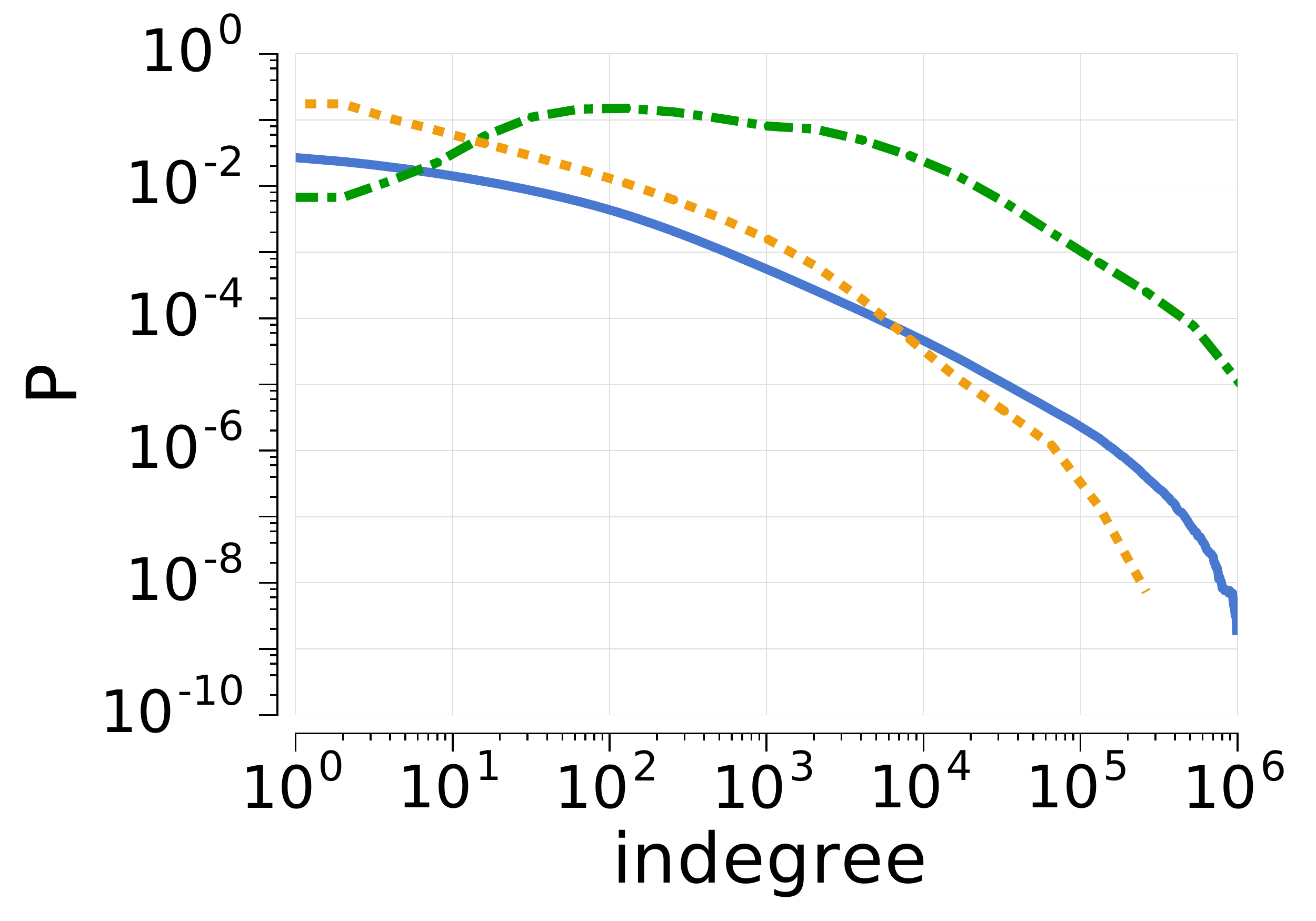}
\includegraphics[width=0.49\columnwidth]{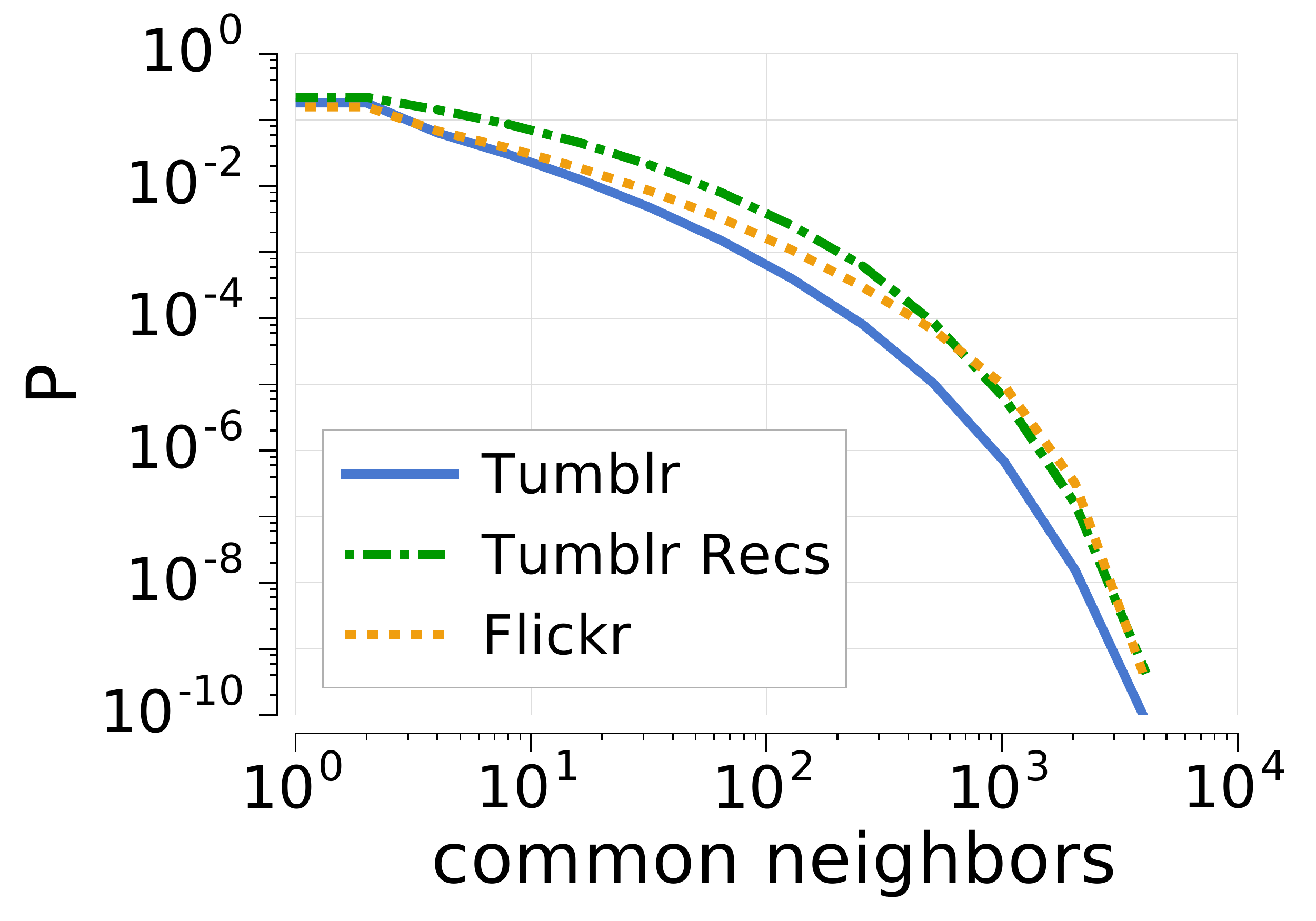}\\
\caption{Distribution of nodes' popularity (indegree) and similarity with ego (common neighbors) at the time of their inclusion in the ego-network.}
\label{fig:predictors_distr}
\end{center}
\end{figure}

\begin{figure}[t!]
\begin{center}
\includegraphics[width=0.49\columnwidth]{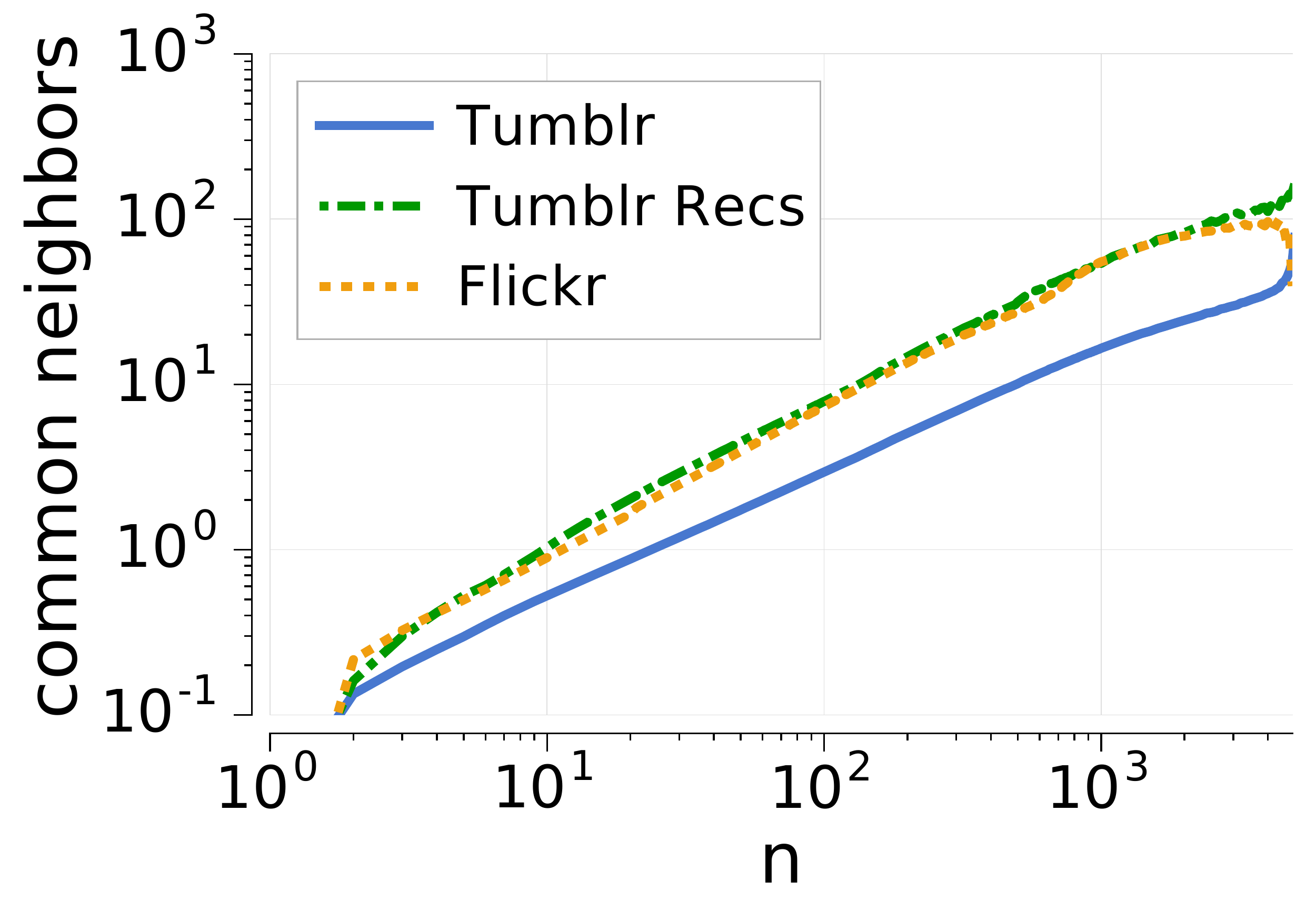}
\includegraphics[width=0.49\columnwidth]{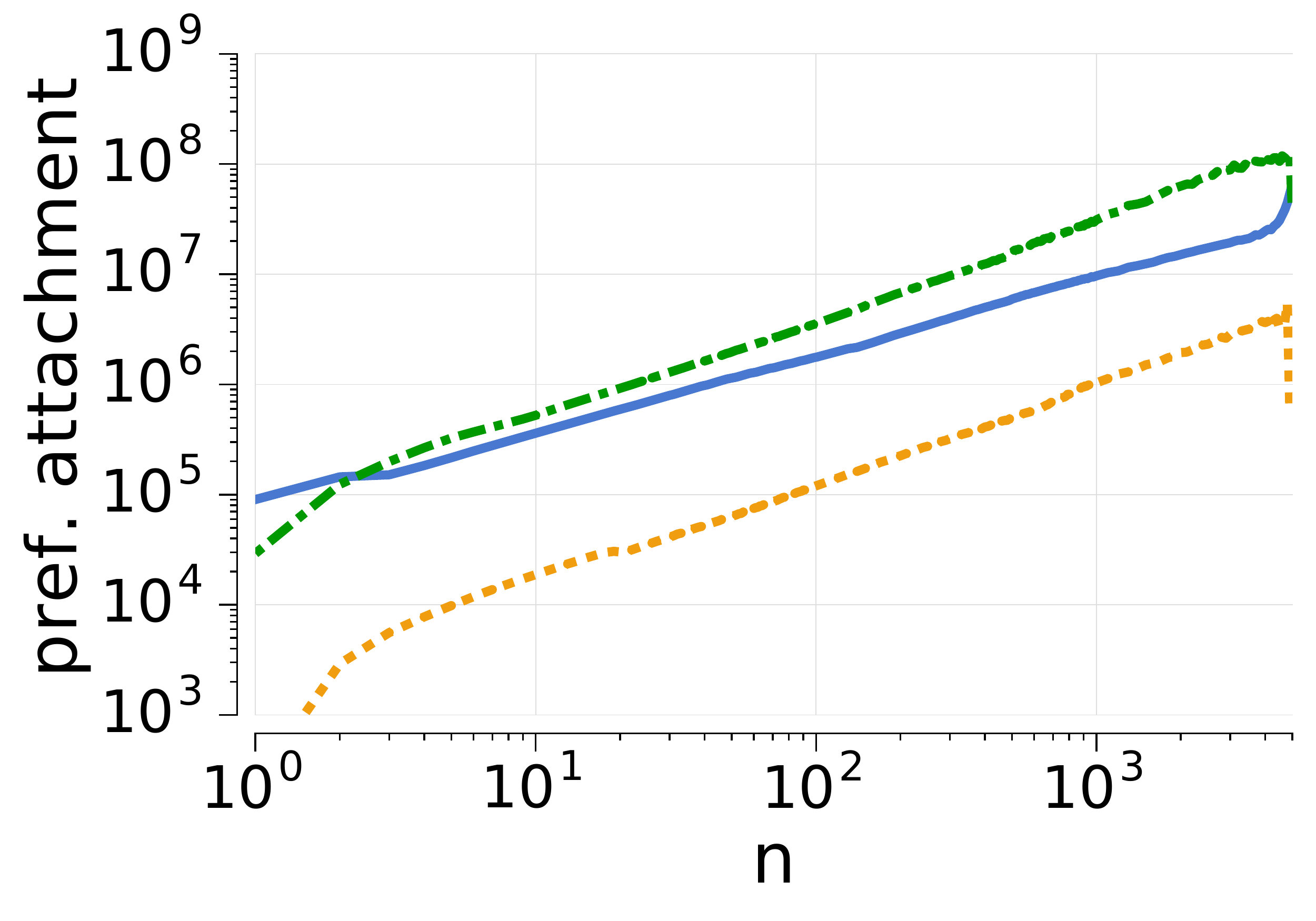}\\
\includegraphics[width=0.49\columnwidth]{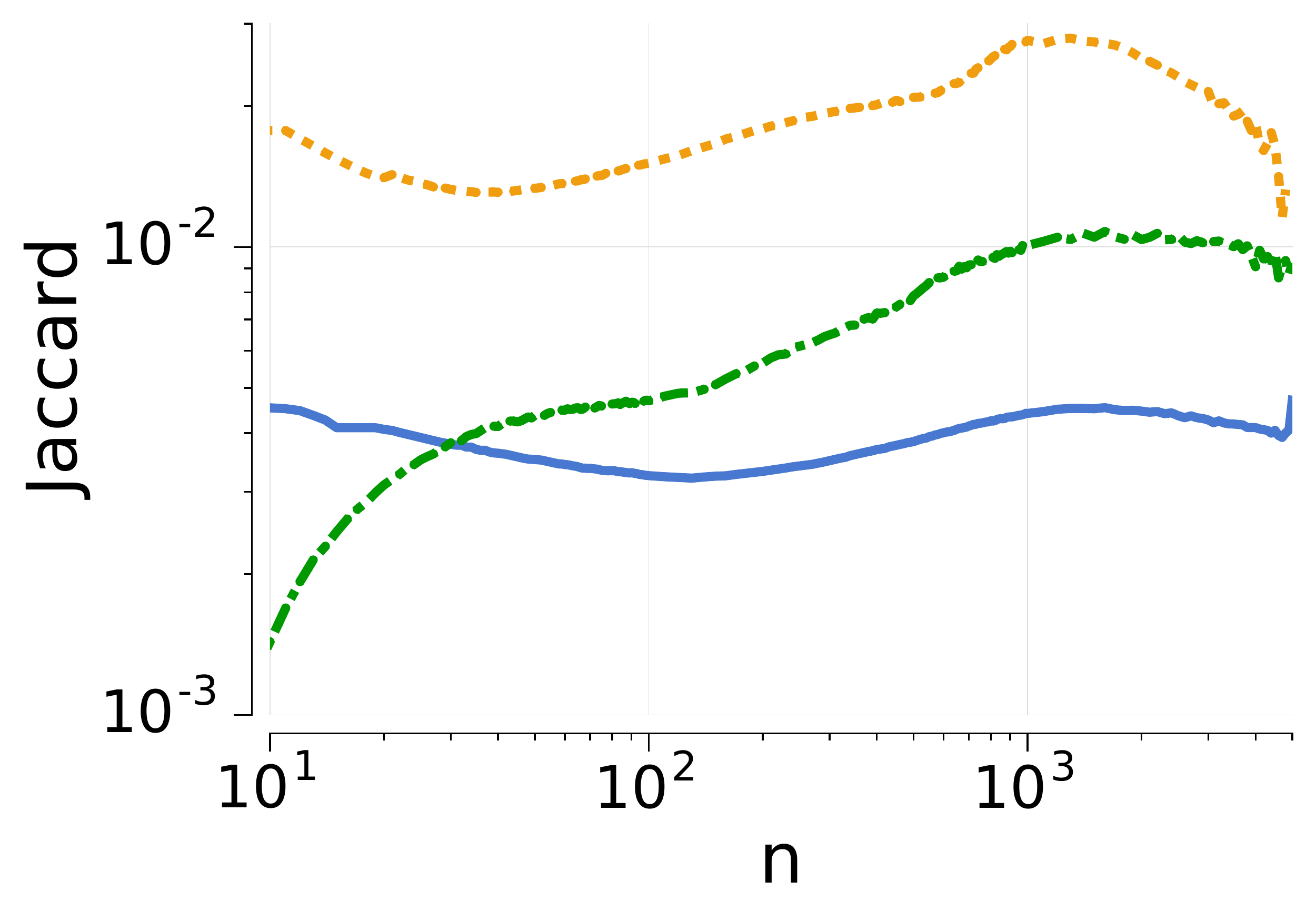}
\includegraphics[width=0.49\columnwidth]{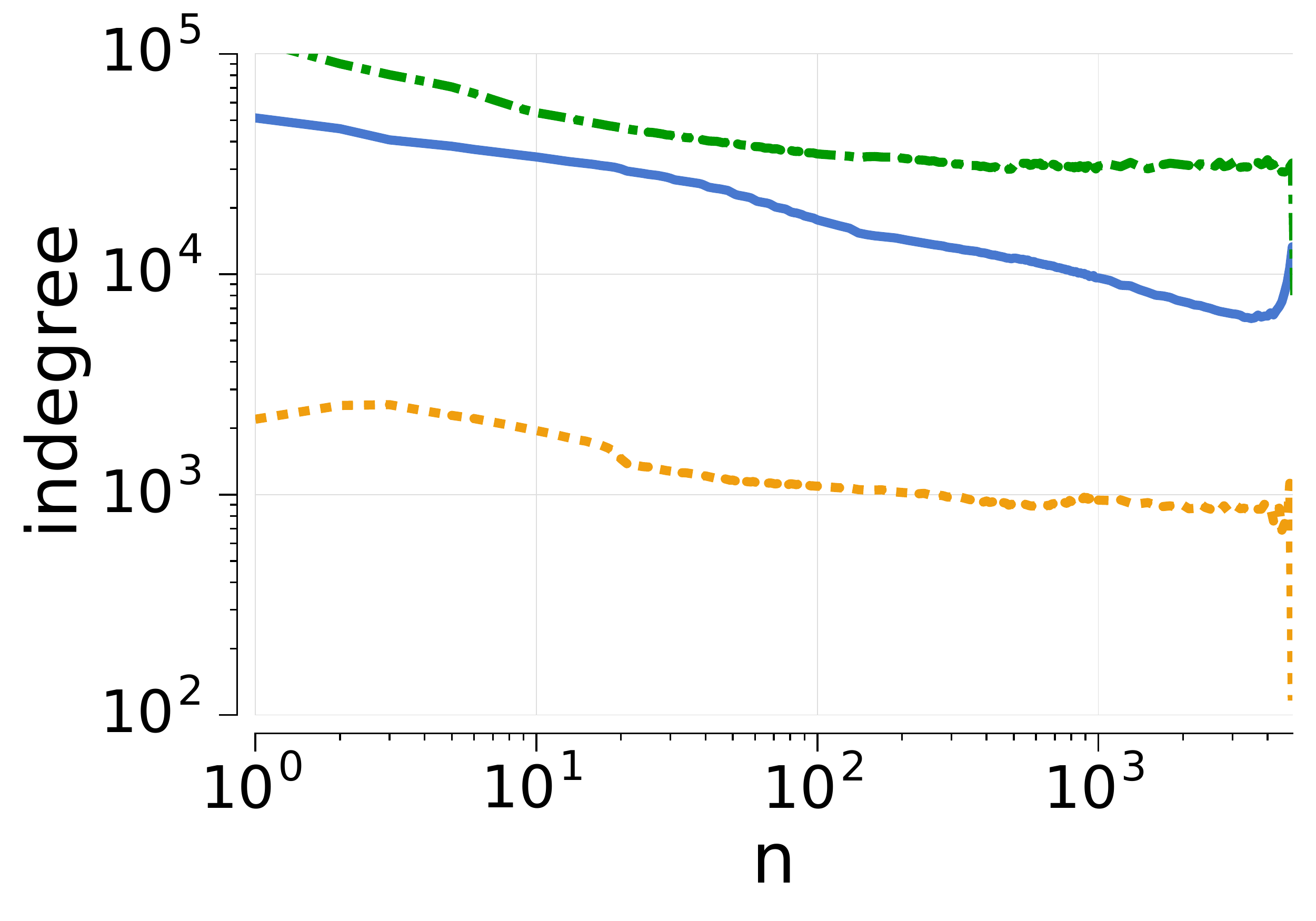}
\caption{Average similarity and popularity indicators of $n^{th}$ node added to the ego-network: common neghbors between the ego and the newly added nodes, Jaccard similarity between their neighbors sets, preferential attachment indicator ($k_{out}(ego) \cdot k_{in}(new node)$), and indegree of the new node.}
\label{fig:predictors_vs_time}
\end{center}
\end{figure}

We select two simple (yet widely-used) proxies of popularity and similarity. Given an ego $i$ who has added $j$ as its neighbor at time $t$, we consider the alter's indegree $k^t_{in}(j)$ as an indicator of its popularity and the number of common neighbors between the ego $i$ and the alter $j$, $CN^t(i,j) = |\Gamma^t_{out}(i) \cap \Gamma^t_{in}(j)|$, as a measure of similarity. Drawing the distributions of $k^t_{in}$ and $CN^t(i,j)$ (Figure~\ref{fig:predictors_distr}), we observe that the range of values is very broad in both platforms. The $CN$ distributions suffer from cut-offs (around $CN=200$) caused by the scarcity of nodes with hundreds of common neighbors or more. Recommended Tumblr nodes yield distributions that are skewed towards higher values because the recommender picks by design those profiles that are popular and well-connected to the ego's neighbors.

As new nodes are added to the ego-network, the number of their common connections with the ego naturally increases (and so does the preferential attachment indicator, as expected). The Jaccard similarity between their neighbor sets oscillates, increasing when the ego-network's size is in the interval $[100,1000]$ and decreasing otherwise. The popularity of new ego-network members, computed as their indegree in the social network, decreases as the ego-network grows. A summary of all the indicators is given in Figure~\ref{fig:predictors_vs_time}. 

All the trends are similar, yet shifted towards higher values, when considering recommended nodes only. In short, recommended nodes tend to share more contacts with the ego and to be more popular, which corroborates previous observations about link recommendations being beneficial mostly to popular nodes~\cite{su16effect}. The indicator that differs the most is the Jaccard similarity, that increases monotonically with $n$ for recommended contacts.

\subsection{Temporal activity} \label{sec:analysis:batches}

\begin{figure}[t!]
\begin{center}
\includegraphics[width=0.49\columnwidth]{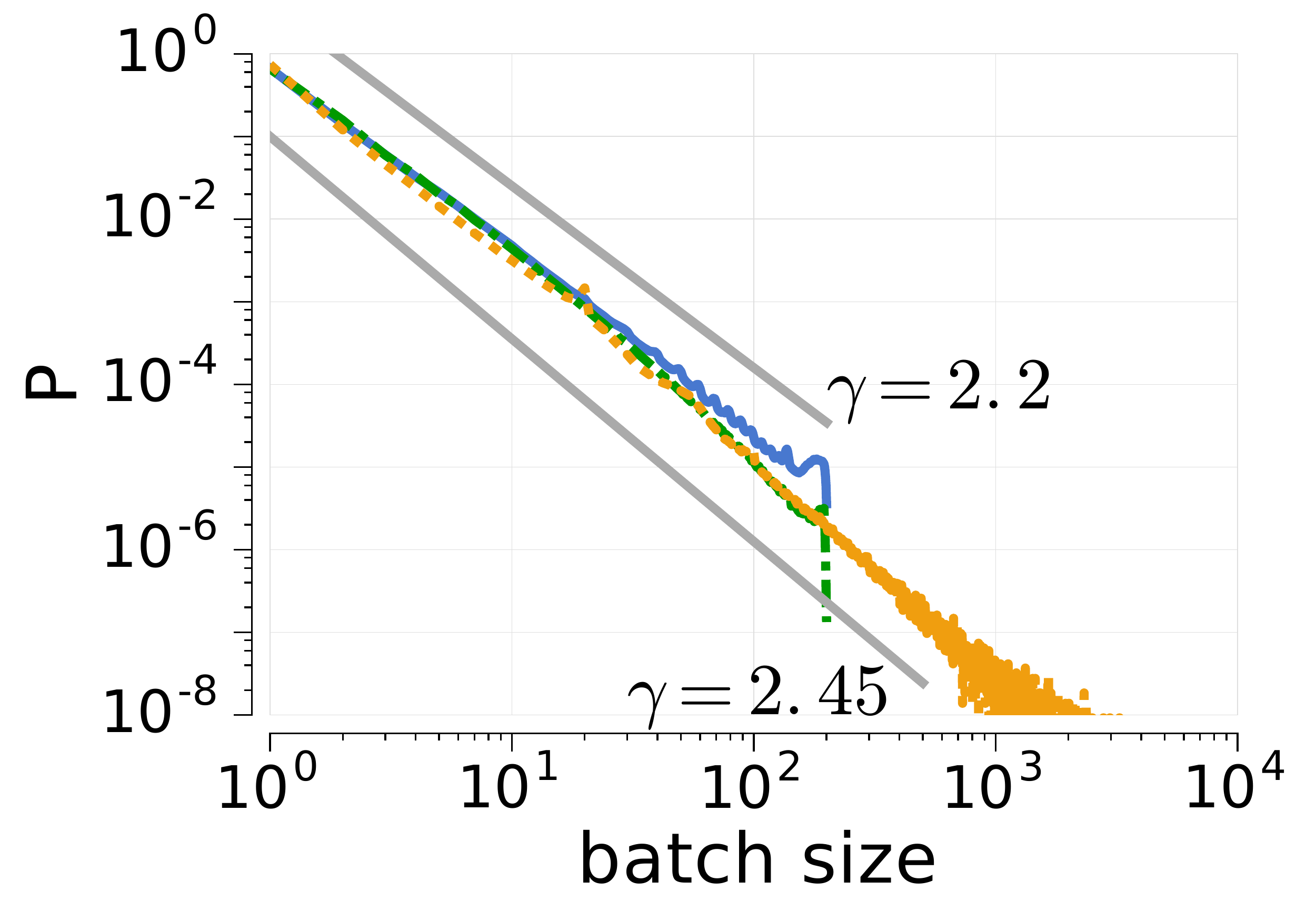}
\includegraphics[width=0.49\columnwidth]{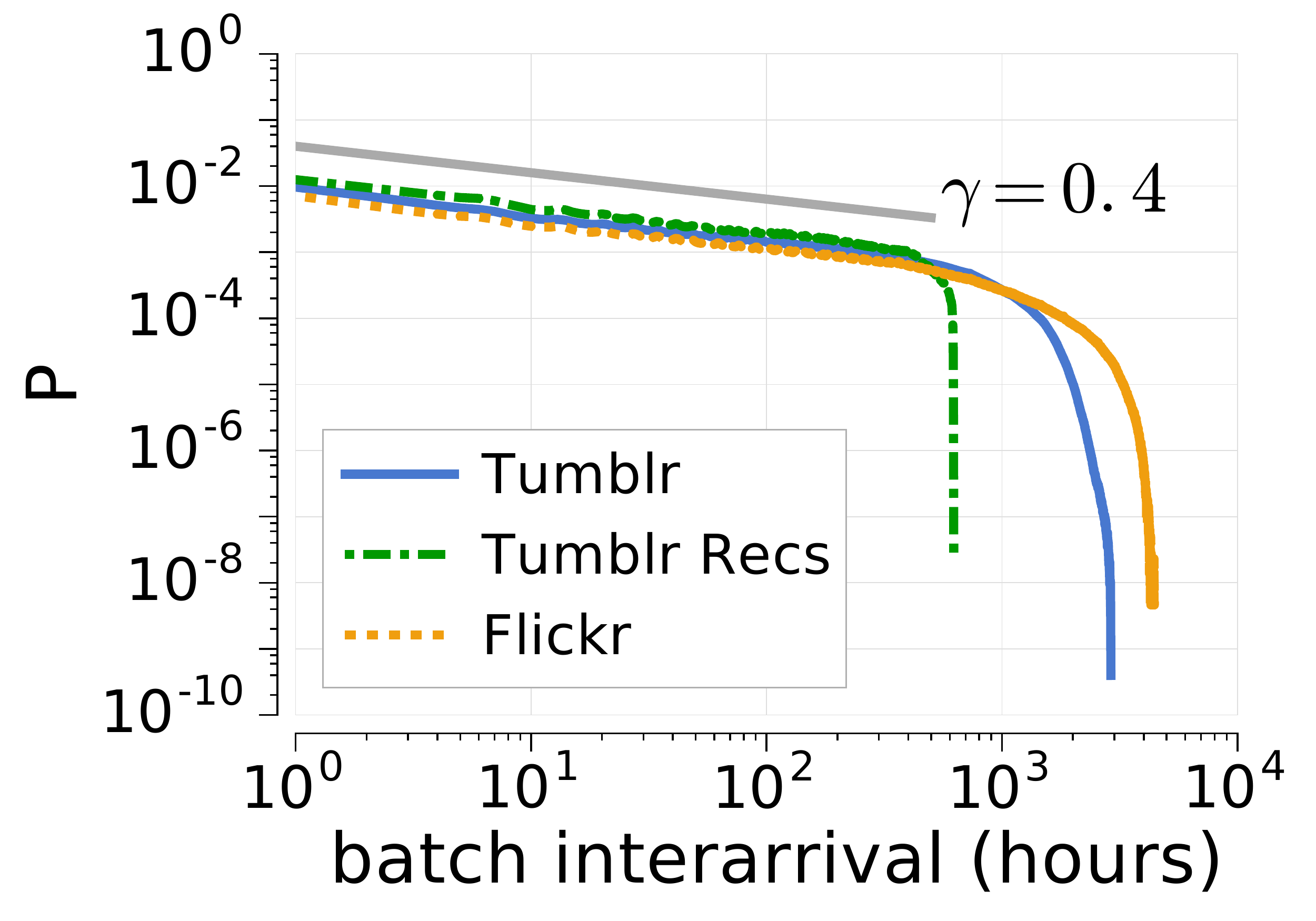}
\caption{Distributions of batch size ($s_b$) and batch interarrival time ($\tau_b$). Best fitting power law exponents reported as reference.}
\label{fig:batches_distr}
\end{center}
\end{figure}

\begin{figure}[t!]
\begin{center}
\includegraphics[width=0.49\columnwidth]{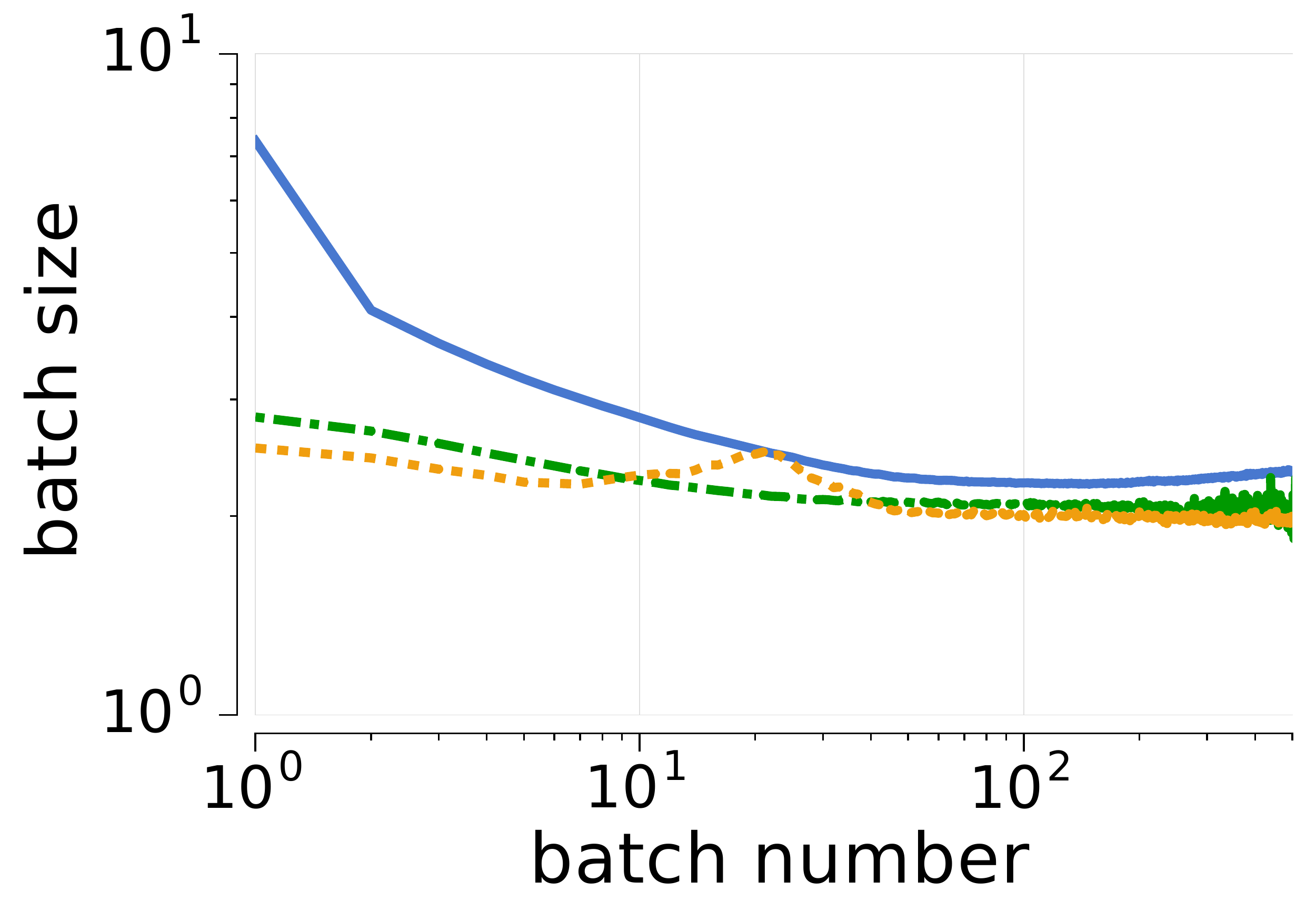}
\includegraphics[width=0.49\columnwidth]{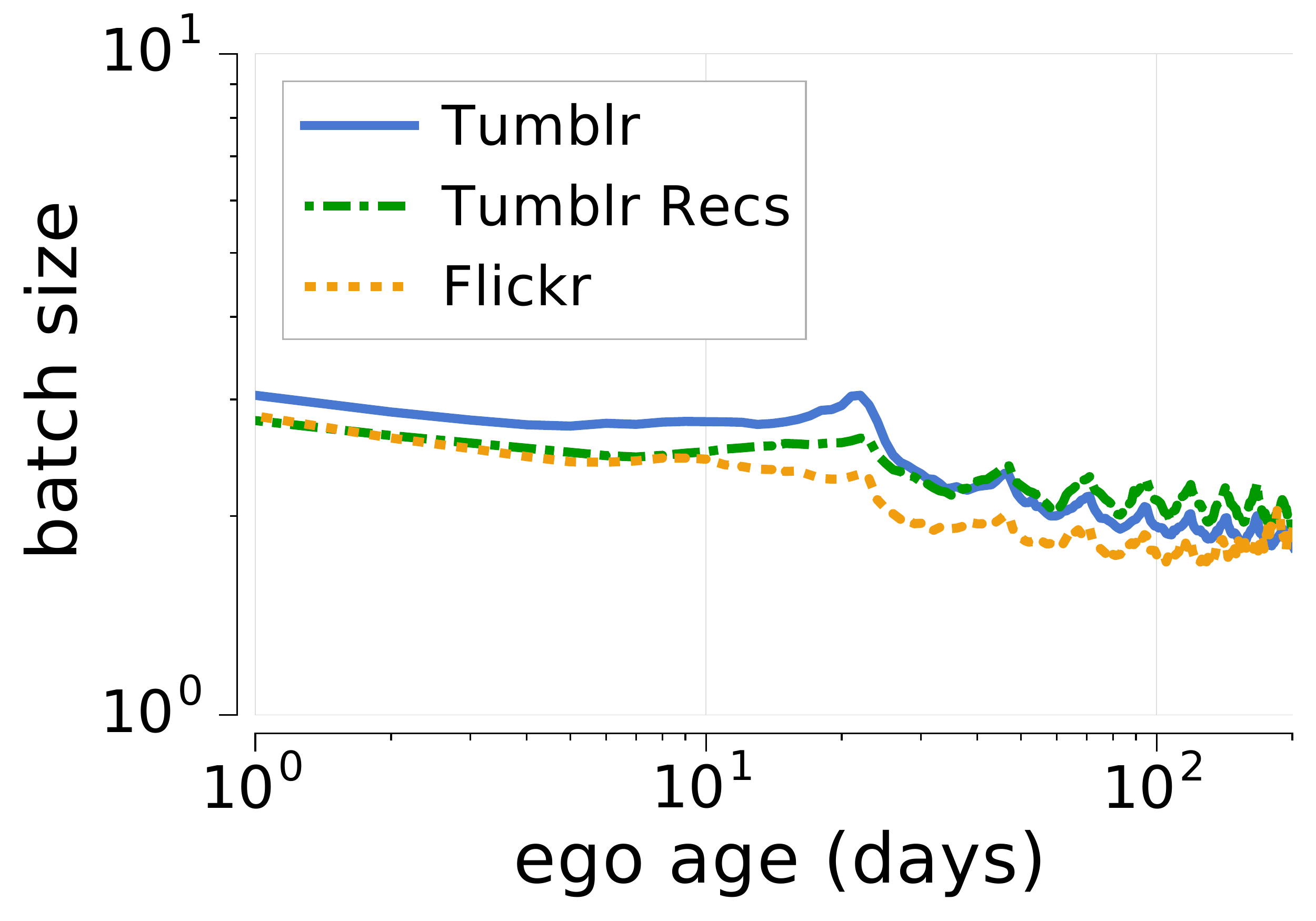}
\caption{Batch size as a function of time, measured as the number of batches created or as the ego's age measured in number of days. The Tumblr Recs curves summarize the trend for batches containing at least one recommended node.}
\label{fig:batches_vs_time}
\end{center}
\end{figure}

Creation of links is not uniform in time. Previous literature found evidence that, globally, the creation of links happens in bursts~\cite{kumar03bursty,kikas13bursty}. At a local level, we aim to learn how often and in which phases of the ego-network life users select new neighbors.

\vspace{4pt}
\noindent \textbf{Q3: When do ego-networks expand?}

To measure how much node additions to an ego-network are concentrated in short periods of time, we resort to basic session analysis to group together temporally-contiguous events. As is standard practice in the analysis of browsing behaviour~\cite{spink06multitasking}, we split user sessions by timeout: a session starts when a new node is added to the ego-network and ends when no other node has been added for 25 minutes. We call \textit{batch} a set of nodes added in a single session. 

We compute the average \textit{batch size} $s_b$ and the session \textit{interarrival time} $\tau_b$, namely the time (hours) elapsed from the session's end to the next session's start. The process of batch creation is bursty when $i)$ there are strong temporal heterogeneities in the interarrival time, and $ii)$ consecutive link creations are not independent events. A standard practice to assess those conditions is to measure the decay of the probability density functions for $s_b$ and $\tau_b$: power law decays in the form $P(x) \sim x^{-\gamma}$ indicate burstiness~\cite{karsai12universal}. The distribution of batch size $s_b$ follows a power-law trend, with exponents $2.2$ and $2.45$ in Tumblr and Flickr, respectively (Figure~\ref{fig:batches_distr} left). In Flickr, the size scales freely as there are no boundaries preventing the addition of any number of contact. In Tumblr, we observe a sharp cutoff at $200$ as the service policy enforces a maximum limit of 200 link creations per user per day. The decay of $\tau_b$ is similar on both platforms, with initial intervals fitting a power-law with exponent $\gamma=0.4$, followed by exponential cutoffs due to the finite time window (Figure~\ref{fig:batches_distr}, right).

\begin{figure}[t!]
\begin{center}
\includegraphics[width=0.90\columnwidth]{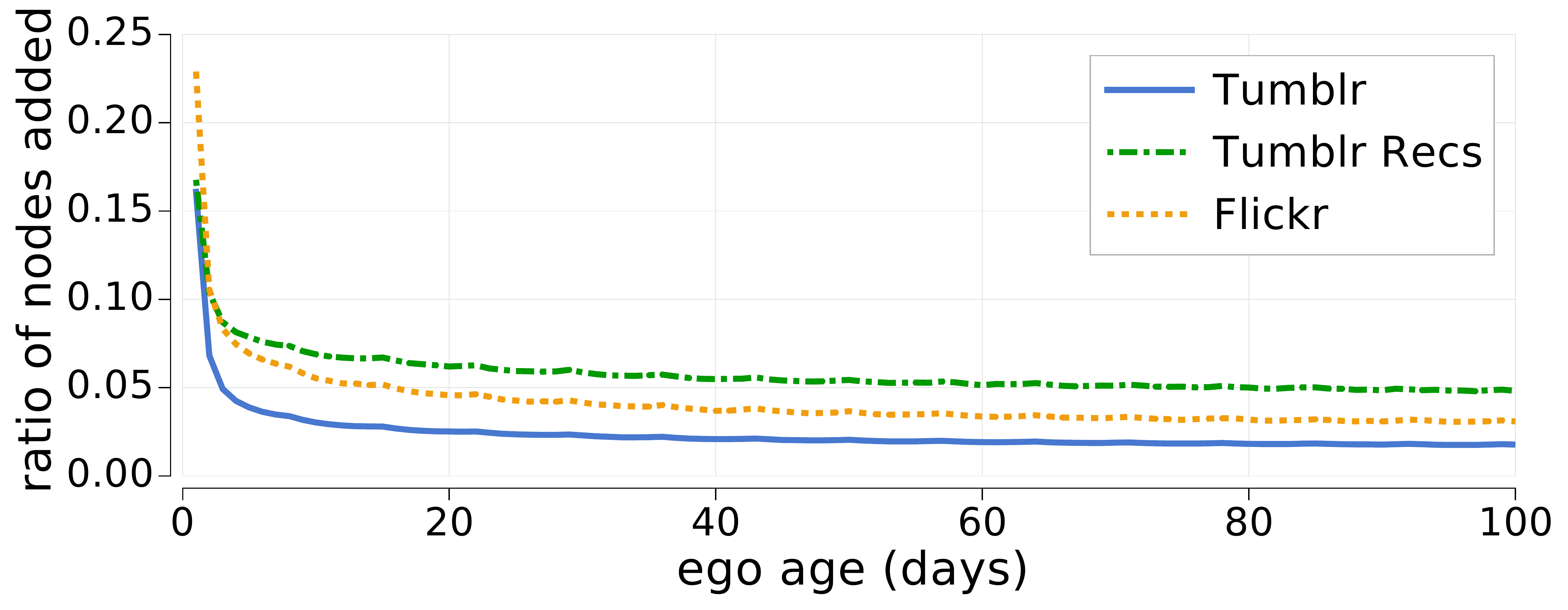}
\caption{Average portion of links created in the first 100 days of the ego's life, relative to the final ego-network size. Only nodes who have created links for at least 6 months are considered.}
\label{fig:link_perc_vs_days}
\vspace{-8pt}
\end{center}
\end{figure}

If we consider only sequences of batches containing at least one recommended link, we see that recommendations are associated with the creation of less links per session, but at higher rate. The average batch size is $2.18$ in Flickr and $2.67$ in Tumblr; Tumblr batches with recommended links are $12\%$ smaller (average size $2.35$). The median interarrival time is relatively high in both platforms ---12 days in Tumblr, 2 weeks in Flickr--- but only 5 days for pairs of consecutive batches containing recommended links. No causal claim connecting recommendations and rate of link creation can be made, as a number of confounding factors could influence this trend (e.g., users who are more active might more naturally engage in recommendations). However, this result provides partial evidence that recommendations might contribute to alter the natural time scale of link creation.

The average batch size decreases as time passes and the ego-network grows (Figure~\ref{fig:batches_vs_time}). After the first 30 days (or the first 20-30 batches created), the batch size stabilizes around 2. A similar decreasing trend is also found for the interarrival time $\tau_b$ (not shown). This suggest that nodes tend to build most of their ego-network in the first stages of their life. To confirm that, we compute the average daily ratio of the total number of the ego-network's nodes added to in the first 100 days of the ego's life. We only consider users whose link creation activity spans at least 6 months, to avoid biases introduced by users with short lifespan. As expected, a big chunk of nodes are typically added in the first days of activity (Figure~\ref{fig:link_perc_vs_days}). This finding adds nuance to previous work on temporal graphs. Studies on Flickr using a coarser temporal granularity found that the raw number of new links created by the ego in time is uniform over time~\cite{leskovec08microscopic}. Here we find that the uniform trend starts only after an initial spike of link creations.

\subsection{Community formation} \label{sec:analysis:community}

Ego-networks have a clear community structure because people tend to interact with multiple social circles (e.g., school friends, family members) that are typically weakly connected to one another~\cite{mcauley14discovering}. We ask about the role of these communities in the graph evolution.

\vspace{4pt}
\noindent \textbf{Q4: Is the ego-network growth driven by the boundaries of its communities?}

The ego-network may follow a depth-first expansion pattern with respect to communties, in which the ego preferentially connects to nodes belonging to a community before exploring others. In Flickr, for example, a person could first follow all the accounts of family members and then those of a photography club. Alternatively, the ego-network may expand either breadth-first, picking new nodes in a round-robin fashion, or regardless of the community structure. In social network analysis research we know little evidence in support of any of these scenarios. In the context of web navigation and search, in-depth exploration of content is often most effective and cognitively more natural~\cite{debra94information,tauscher97how}; we hypothesize that the same holds for community exploration.

\begin{figure}[t!]
\begin{center}
\includegraphics[width=0.49\columnwidth]{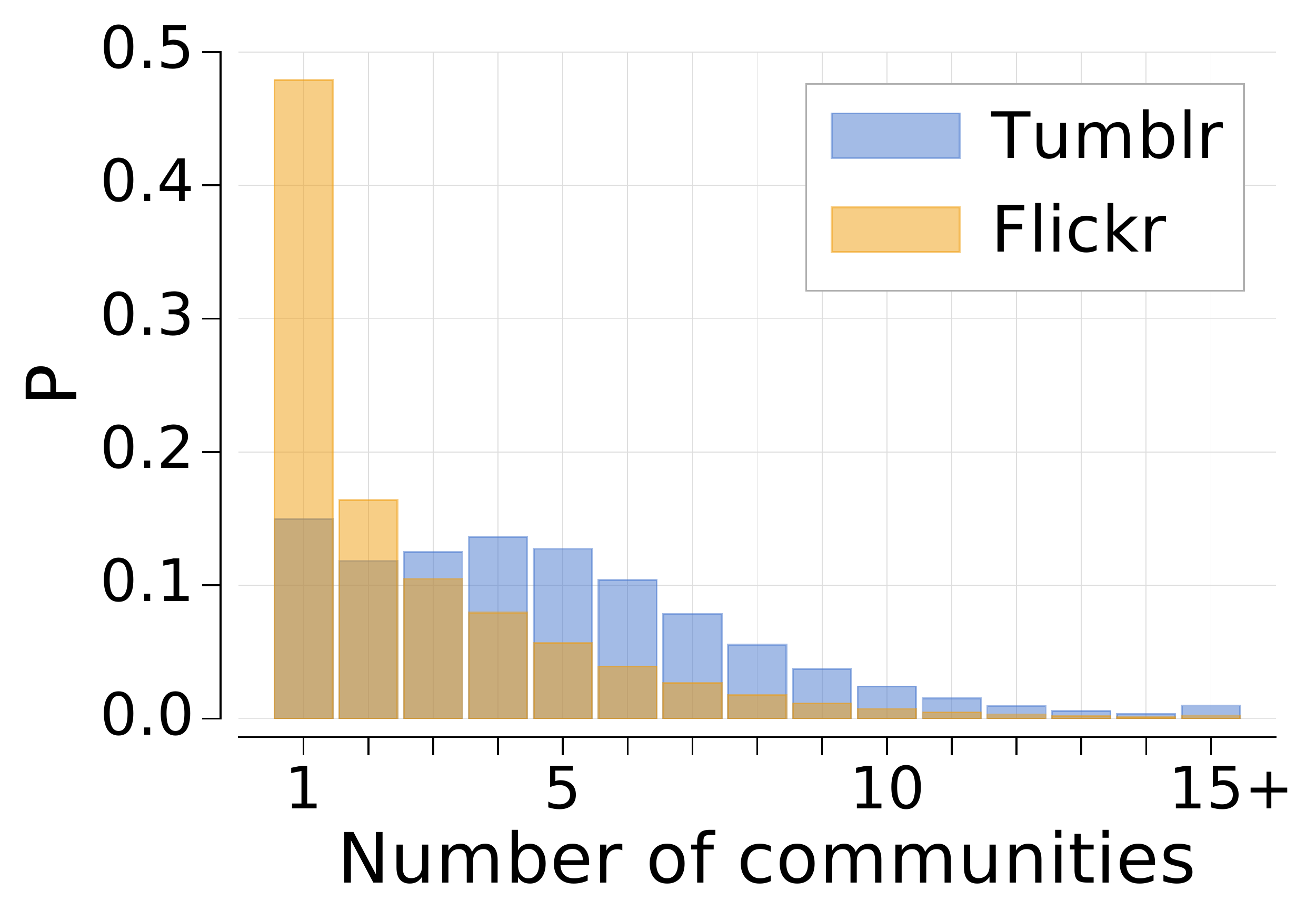}
\includegraphics[width=0.49\columnwidth]{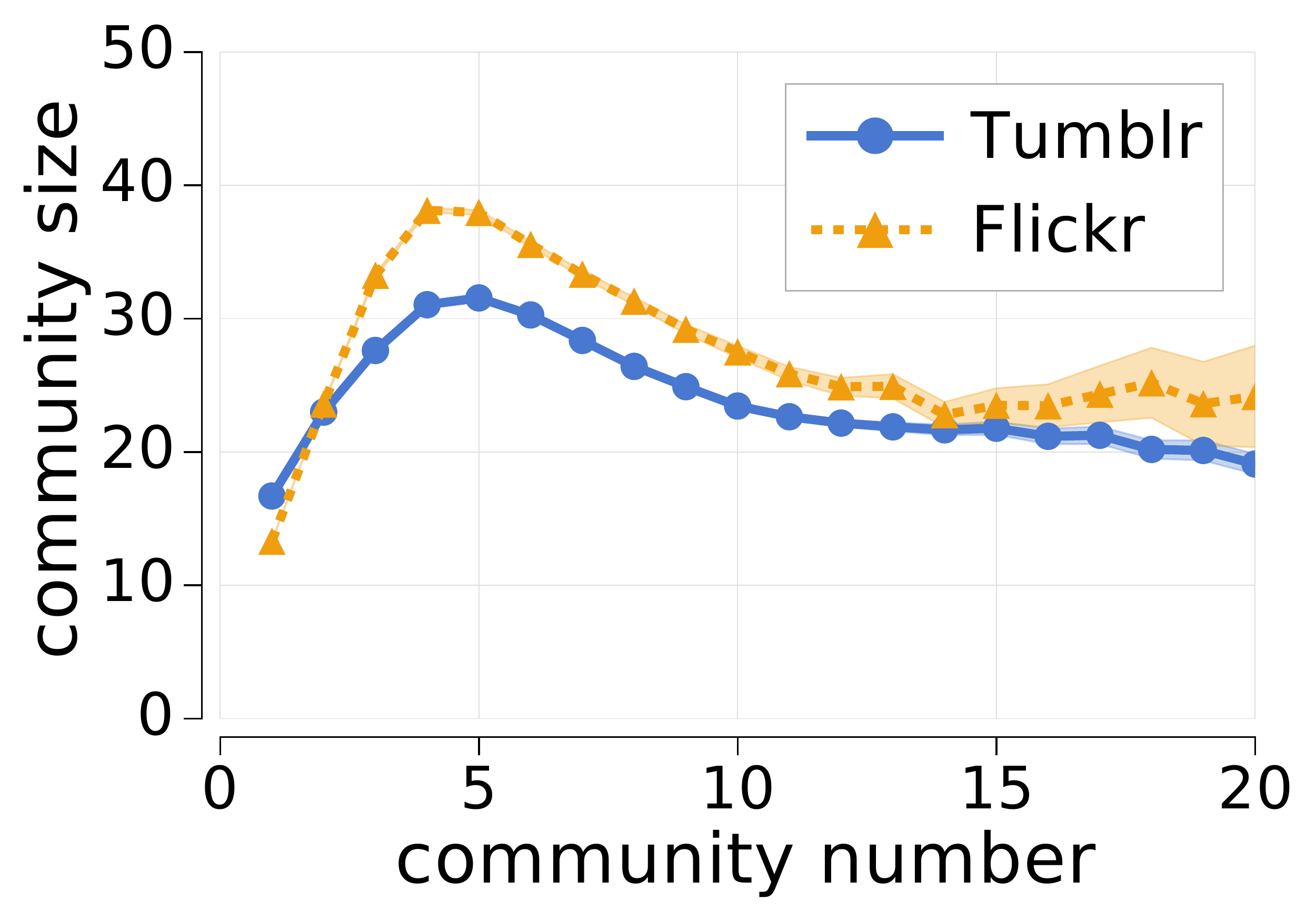}
\caption{Left: distribution of number of communities in ego-networks. Right: average size of communities as they appear in the ego-network; 95\% confidence intervals are shown.}
\label{fig:community_no_distr}
\vspace{-8pt}
\end{center}
\end{figure}

Measuring the extent to which any of these scenarios reflects people's behaviour is challenging, as in reality communities can be overlapping and change their boundaries as the graph grows. Leaving more advanced measurements for future work, we assume a static, hard partitioning of nodes in communities. For all ego-networks, we compute non-overlapping communities\footnote{\small We use the community detection algorithm by Waltman and Eck~\cite{waltman13smart} that is an optimization of the Louvain method~\cite{blondel08fast}.} at time $t=T_{end}$ (the most recent snapshot in our data), heuristically filtering out small ego-networks with less than 5 members. Ego-networks are often composed by few communities, rarely more than 10 (Figure~\ref{fig:community_no_distr}, left).

\begin{figure}[t!]
\begin{center}
\includegraphics[width=0.49\columnwidth]{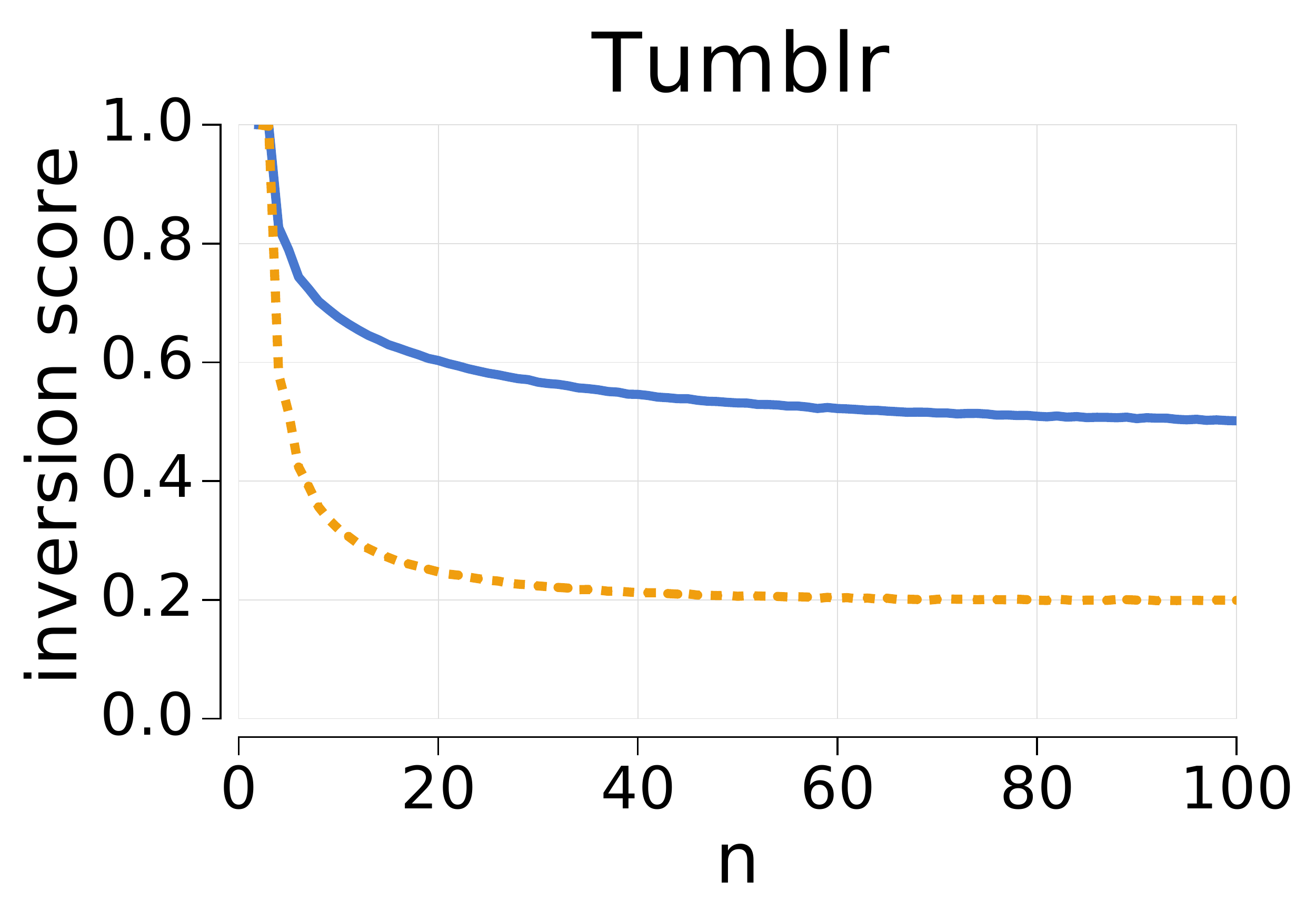}
\includegraphics[width=0.49\columnwidth]{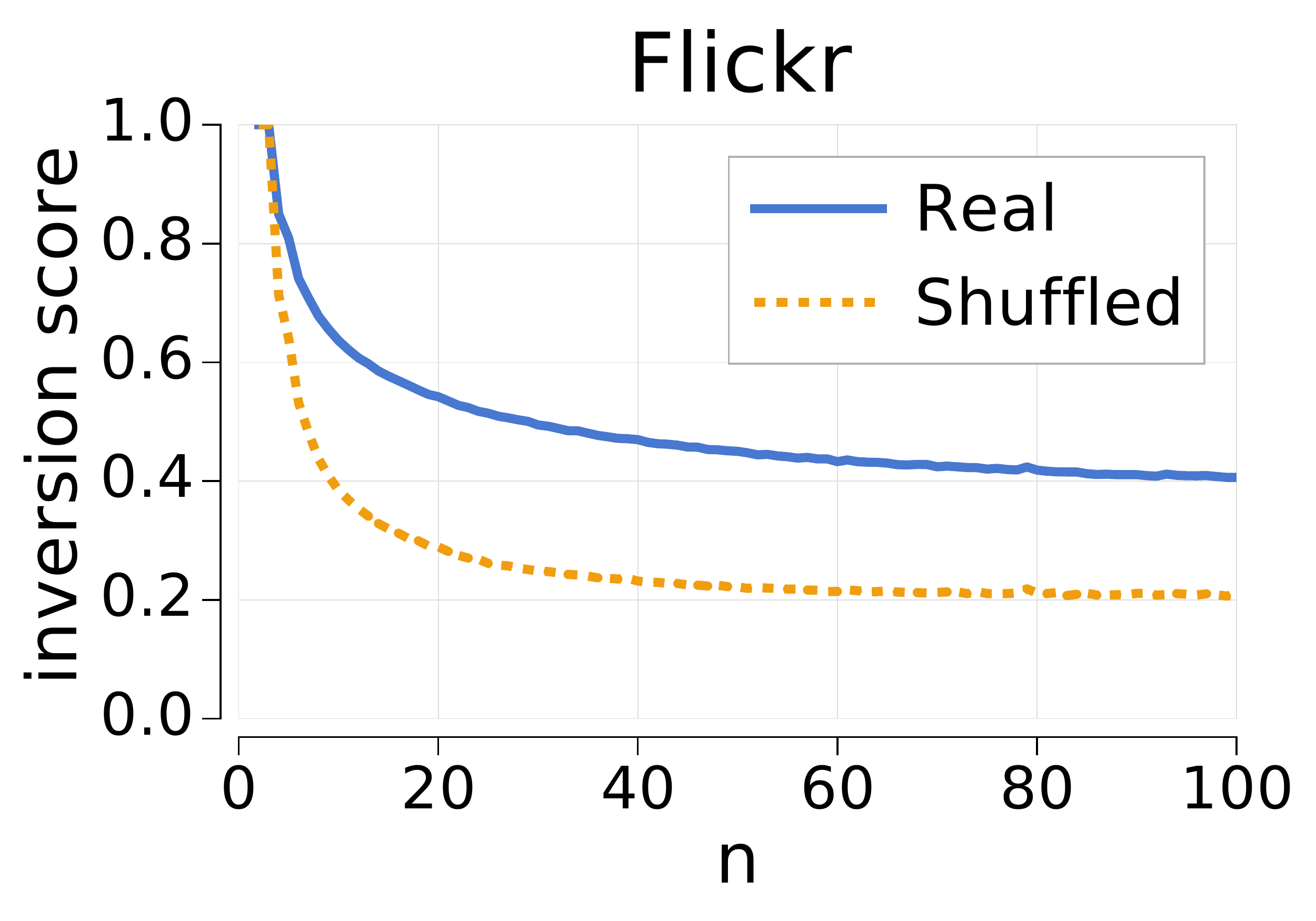}
\caption{Average inversion score of ego-network communities as new nodes are added, compared to a randomized null-model.}
\label{fig:community_inversions}
\end{center}
\end{figure}

To assess to what extent communities emerge over time in an orderly fashion, we rank them by the time the ego has created connections with their nodes. Specifically, we first sort all the nodes by the time they are added to the ego-network (e.g., $[j_1,j_2,j_3,j_4]$). We then replace nodes with the communities they belong to (e.g., $[c_1,c_1,c_2,c_1]$), rank communities by the median position $p$ of their occurrences in that vector (e.g., $p(c_1)=2, p(c_2)=3$; $c_1$ is ranked first, $c_2$ is ranked last), and finally replacing the communities with their respective ranks, thus obtaining a sequence of community ranks $\mathcal{R}$ (e.g., [1,1,2,1]). The intuition is that a community half of whose members has been added to the ego-network by time $t$ comes temporally before any other whose majority of nodes are still outside the ego-network at the same time $t$. If exploration of communities is purely in-depth, $\mathcal{R}$ is fully sorted. The sortedness of a list $L=\{x_1, ..., x_m\}$ can be measured by its \textit{inversion score}:
\begin{equation*}
inv(L) = 1 - \frac{2 \cdot |\{(x_i,x_j) | i < j \wedge x_i > x_j\}|}{\binom{|L|}{2}} \in [-1,1]
\end{equation*}
$inv=1$ indicates sortedness, $inv=-1$ inverse ordering, and $inv=0$ randomness. In Figure~\ref{fig:community_inversions} we plot the average inversion score of $\mathcal{R}$ against the ego-network size. To account for the community size heterogeneity, we compare it with a null-model where the elements in $\mathcal{R}$ are randomly reshuffled. The inversion score quickly stabilizes as the network grows and it has values that are consistently higher (double or more) than the null-model's, supporting the hypothesis that communities tend to be explored in depth, one after the other. 

\begin{figure}[t!]
\begin{center}
\includegraphics[width=0.49\columnwidth]{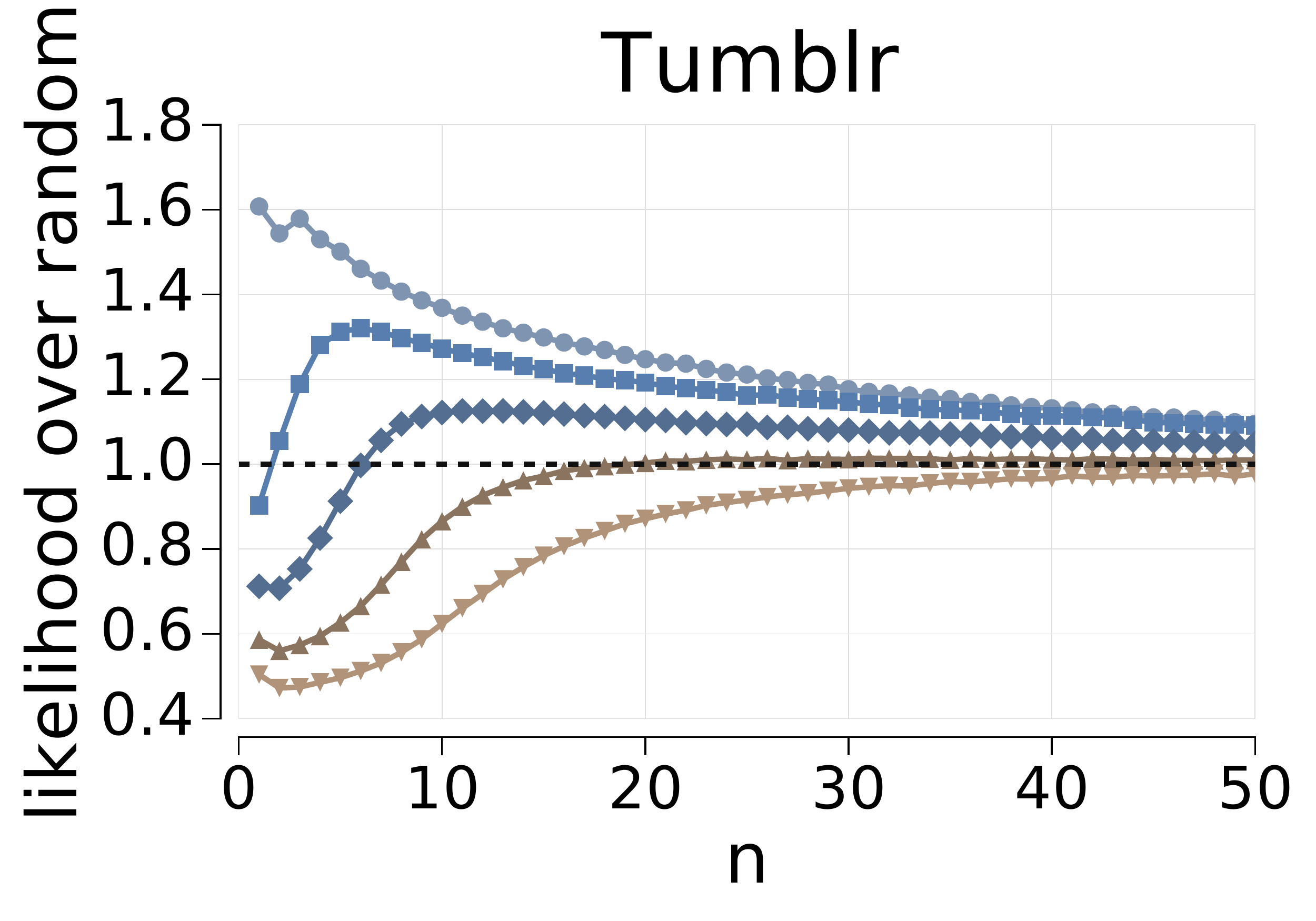}
\includegraphics[width=0.49\columnwidth]{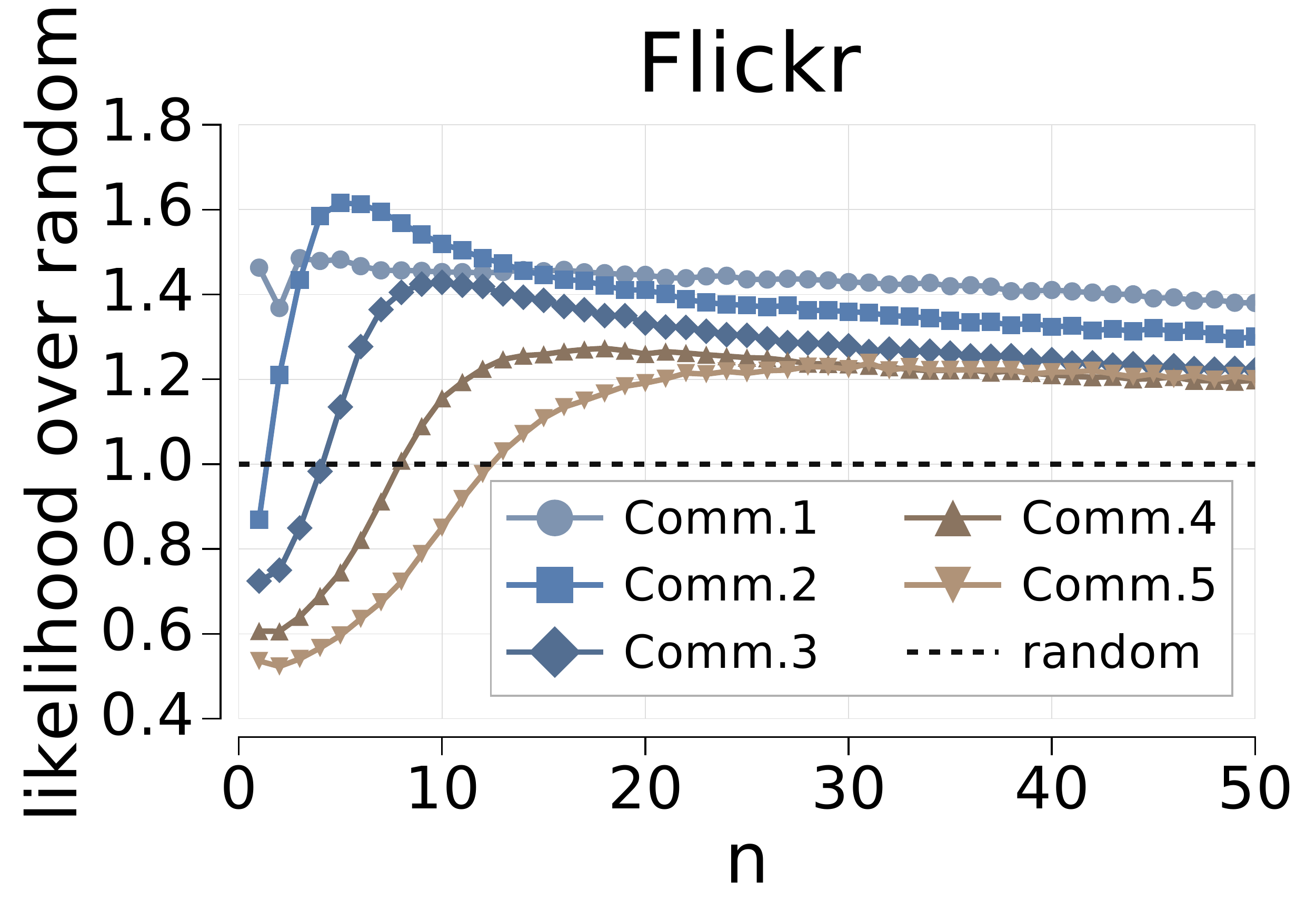}
\caption{Average likelihood (over random chance) that the $n^{th}$ node in a ego-network belongs to the $k^{th}$ community.}
\label{fig:communities_vs_time}
\vspace{-15pt}
\end{center}
\end{figure}

Further evidence can be provided by measuring the probability that the $n^{th}$ node added to the ego-network belongs to the $k^{th}$ community in the ranking, normalized by the probability in the null-model; values higher than 1 indicate above-chance likelihood of a node being in a given community. Figure~\ref{fig:communities_vs_time} shows the average normalized likelihood of the first 50 nodes to belong to the first 5 communities in the rank. Curves for increasing values of $k$ emerge above the randomness threshold one after the other, which backs the hypothesis of communities being explored in-depth.

The size of a community (measured at time $T_{end}$) varies with its temporal rank (Figure~\ref{fig:community_no_distr}, right). On average, people create increasingly larger communities up to the fifth one; from the sixth one on, new communities added become smaller and smaller.

\subsection{Recommendations diversity} \label{sec:analysis:matching}

Our analysis shows that the statistical properties of recommended links are different from those of spontaneous ones. It is known that recommender systems postively affect user engagement, in terms of time spent, content consumption, and user contribution~\cite{freyne09increasing}. In agreement with established knowledge, we have found that users exposed to recommendations create more links, more frequently. It is harder to assess whether recommendations foster or limit access to diverse types of content. The academic  debate about recommendations being the bane or boon of social media is still very lively~\cite{pariser12filter,bakshy15exposure,sharma15estimating}, with evidence brought in support of the two views. We aim to provide further evidence to shed light on this point in the context of link recommenders.

\vspace{4pt}
\noindent \textbf{Q5: Do link recommendations foster diversity?}

It is hard to infer causality from observational data. Matching is a statistical technique that is used to evaluate the effect of a treatment on a dependent variable by comparing individuals who have received the treatment with others with similar observable features who did not receive it. The more similar the paired individuals and the higher the number of pairs, the higher the confidence of estimating the cause of the treatment on the dependent variable. 

\begin{figure}[t!]
\begin{center}
\includegraphics[width=0.75\columnwidth]{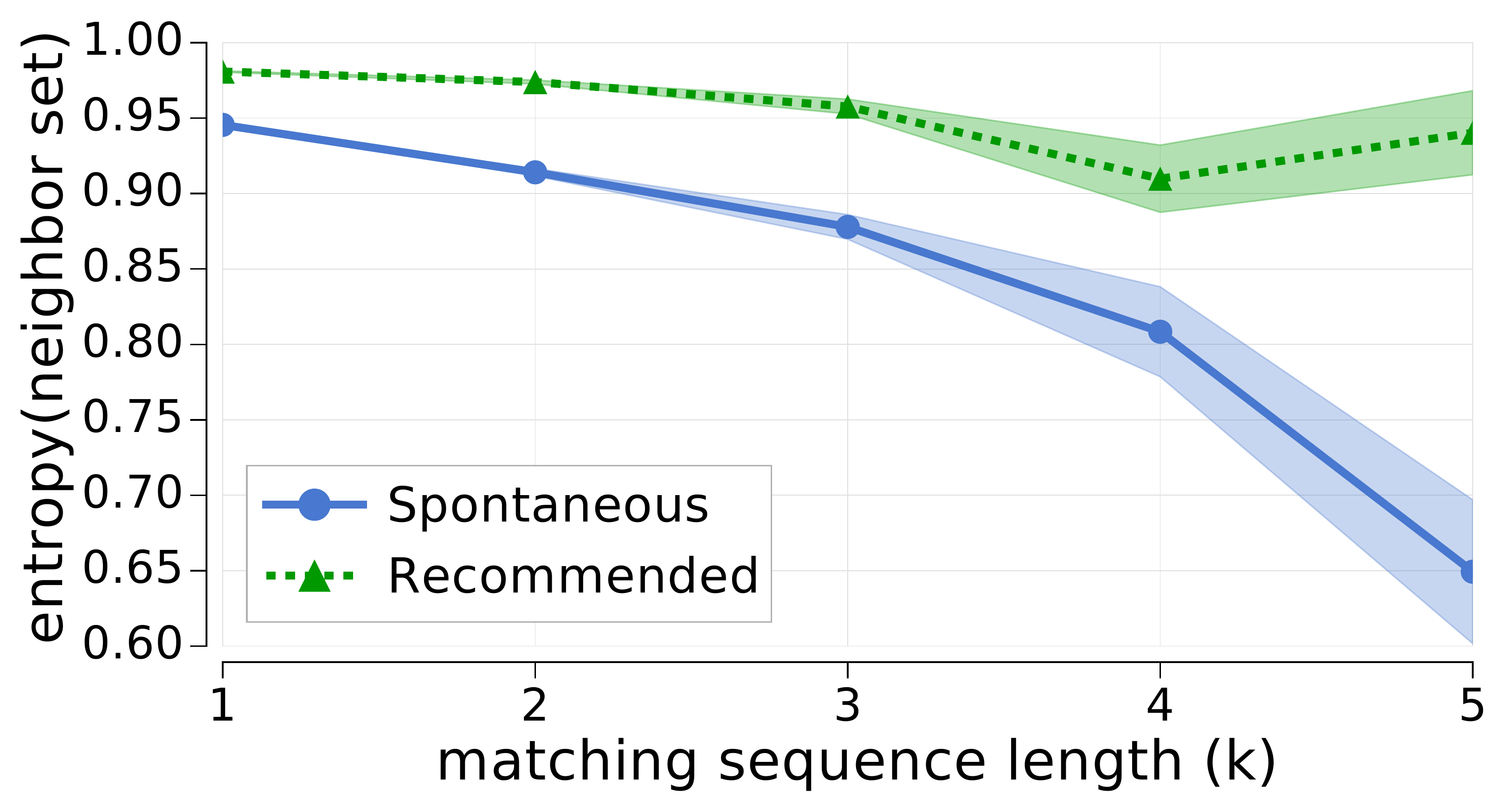}
\caption{Matching experiment. Average entropy of the neighbor sets at step $k+1$ for groups sharing a matching sequence of length $k$. 95\% confidence intervals are shown.}
\label{fig:matching}
\end{center}
\end{figure}

We conduct a matching experiment to measure if people who follow recommendations end up having ego-networks more similar to one another than if they were to ignore recommendations. We arrange people in \textit{matching groups} containing users who are nearly identical in terms of their local connectivity. We assign to the same matching group users who $i)$ registered to Tumblr less than 30 days apart, and $ii)$ whose very first $k$ neighbors are the same and have been added in the same order to their ego-networks. Each matching group is then split in two subgroups: a \textit{treatment} group of users whose $k+1^{st}$ contact is a recommended one, and a \textit{control} group of all the remaining users, whose $k+1^{st}$ contact has been created spontaneously. The variety of contacts created at step $k+1$ is our dependent variable. If the variety measured in one group is significantly different from the other, we can attribute the divergence to the effect of the recommendation, as the initial conditions of the two groups are virtually identical. For example, if we found that the variety of nodes in the treatment group is lower than the one measured on the control group, we would conclude that recommendations conform the process of link creation by inducing users to follow a more restricted set of accounts compared to what would happen by spontaneous user behavior. We measure diversity of contacts through their entropy. To ensure a fair comparison that accounts for size heterogeneity, we use normalized entropy $\widehat{H}$. Given a bag of nodes $X$ of size $N$, where $p(x)$ is the number of occurrences of node $x \in X$ divided by $N$, the normalized entropy is defined as:
\begin{equation*}
\widehat{H}(X) = \sum_{x \in X} \frac{p(x) \cdot log_2(p(x))}{log_2(N)} \in [0,1].
\end{equation*}

Figure~\ref{fig:matching} shows the results for a total of approximately $25K$ matching groups, for $k \in [1,5]$; requiring $k$ identical links is a too strong requirement for larger $k$. Every point is the average of all the matching groups for a given $k$. First, we observe that, the higher the $k$, the lower the entropy. That is expected: the higher the number of common neighbors, the more likely the next selected neighbor will be the same. Last, most importantly, the variety of the treatment group is always higher; this indicates that recommendations foster diversity. Although it is difficult to pin down the exact reasons why this happens, we provide a possible interpretation. Even if the list of recommended contacts was the same for all the users in the treatment group, the reshuffling of the top recommended contacts that Tumblr implements in the link recommender widget introduces asymmetries across users. More generally, we could hypothesize that the recommender system exposes users to a wider set of potential contacts than the ones they would be exposed to by browsing or searching on the site, thus providing a wider spectrum of options and, in turn, a more diverse set of individual choices.

To test the robustness of the results, we explored some possible alternatives in the setup of the matching experiment. Specifically we: $i)$ measured the entropy at $k+2$ instead of $k+1$ (we leave the $k+n$ generalization for future work); $ii)$ selected only control and treatment groups with at least $m \in [2,15]$ members (the results reported are for $m=5$); $iii)$ randomly downsampled the larger group to match the size of the smaller one, to balance the size of the two; $iv)$ run two independent experiments including in the treatment group only users whose recommended contact had 1) at least one common neighbor (i.e., the recommendation is provided based on network topology features) or 2) no common neighbors (i.e., the recommendation is provided on a topical basis). The absolute values vary slightly across setups, but the qualitative results remain the same.

\section{Impact on link prediction} \label{sec:prediction}

\begin{figure}
\begin{floatrow}
\capbtabbox
{
\begin{tabular}{l|cc}
\hline
\multicolumn{1}{c}{\textsf{Model}} & \multicolumn{1}{c}{\textsf{AUC}} & \multicolumn{1}{c}{\textsf{F-Score}}\\
\hline
Baseline & 0.893 & 0.813 \\
+ age & 0.938 & 0.864 \\
+ $k_{out}$ & 0.897 & 0.817 \\
All & $\mathbf{0.943}$ & $\mathbf{0.87}$ \\
\hline

\multicolumn{1}{c}{} & \multicolumn{1}{c}{} & \multicolumn{1}{c}{}\\
\end{tabular}
}
{\captionsetup{width=.9\linewidth}
  \caption{Link prediction results.}
\label{tab:prediction}
}
\hfill
\hspace{-1cm}
\ffigbox
{
\includegraphics[width=0.40\textwidth]{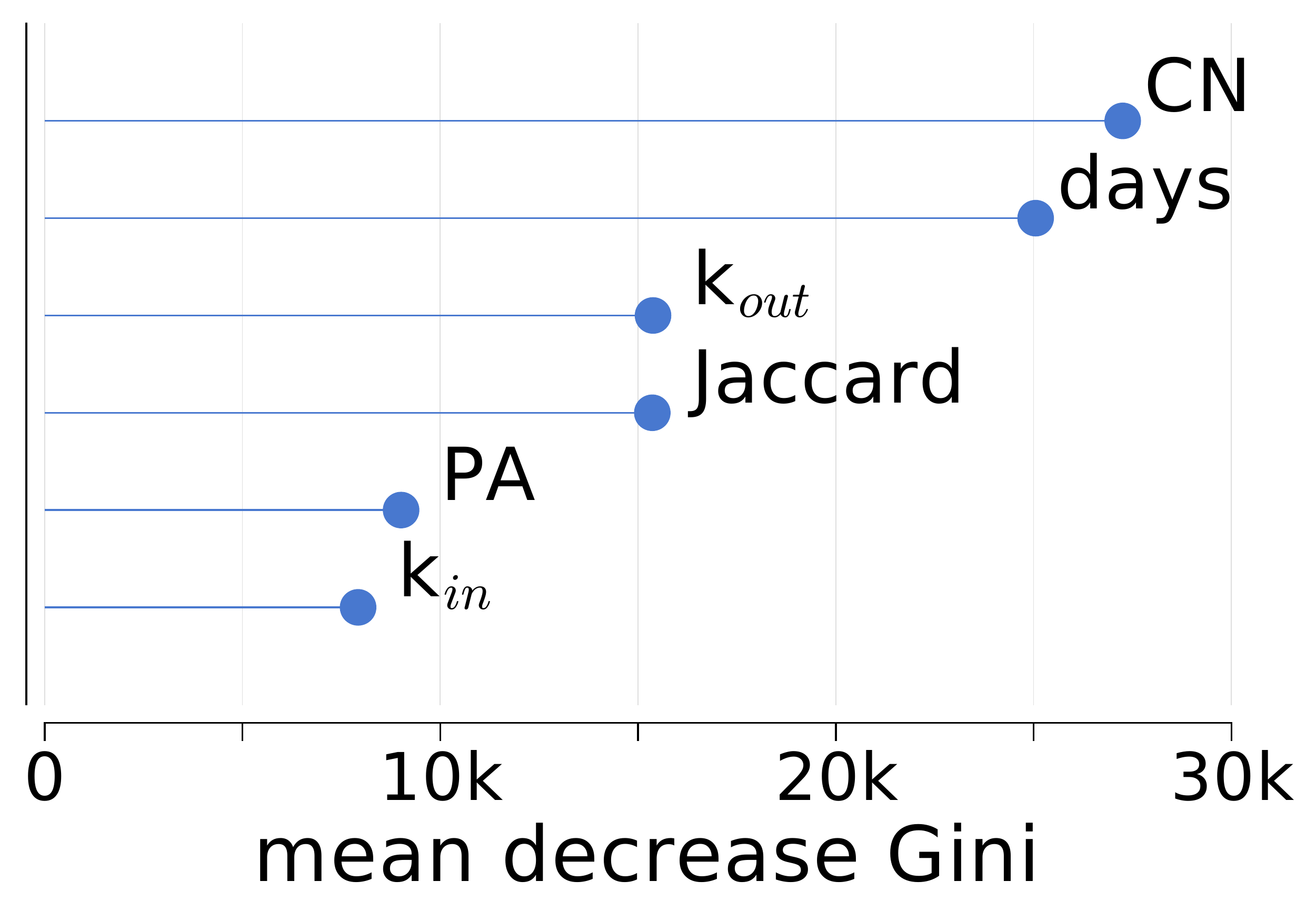}
}
{
\captionsetup{width=.9\linewidth}
  \caption{Feature importance.}
\label{fig:featureImportance}
}
\end{floatrow}
\end{figure}

Our analytical results have direct implications on how link recommender systems can be enhanced to provide more effective suggestions. Next, we discuss two prediction experiments that aim to answer two research questions.

\vspace{4pt}
\noindent \textbf{Q6: To what extent temporal features improve the ability to predict new links?}

The previous analysis showed that egos add new neighbors to their network with criteria that change in time. The resulting hypothesis is that link recommendations that adapt to the current evolutionary stage of the ego-network could gain effectiveness. To test such hypothesis, we run a prediction experiment.

We consider a snapshot of the Tumblr social network at an arbitrary time $t$ (January $1^{st}$ 2015). To build a training set, we sample $200K$ node pairs $(i,j)$ that are not directly connected but with at least one directed common neighbor (i.e., there is a directed path of length 2 from $i$ to $j$). Half of the pairs will be directly connected by a link  from $i$ to $j$ before $T_{end}$ (positive examples), the remaining half will remain disconnected (negative examples). For every pair, we extract six simple features: $i$'s outdegree ($k_{out}(i)$), $j$'s indegree ($k_{in}(j)$), preferential attachment (PA~$= k_{out}(i) \cdot k_{in}(j)$), common neighbors (CN~$ = |\Gamma^{t}_{out}(i) \cap \Gamma^{t}_{in}(j)|$), Jaccard similarity between neighbor sets (Jaccard = $\frac{|\Gamma^{t}_{out}(i) \cap \Gamma^{t}_{in}(j)|}{|\Gamma^{t}_{out}(i) \cup \Gamma^{t}_{in}(j)|}$), and $i$'s age measured in number of days elapsed from $i$'s profile creation to $t$. Age and $i$'s outdegree are the two temporal features whose effectiveness we want to investigate: one measures time on a continuous scale, the other on the discrete scale of link creation events.

The simple features above can predict the node pair class very accurately, and a summary of this evaluation (10-fold cross validation using random forest) is given in Table \ref{tab:prediction}. The model trained on the full set of features yields an AUC of 0.943 and a F-measure of 0.87, which is an improvement of $5.6\%$/$7\%$ in terms of AUC/F-measure over a baseline model that does not consider temporal features. Adding the feature $k_{out}$ to the baseline model yields only a slight improvement in accuracy ($0.45\%$/$0.49\%$), while considering age improves the accuracy in a more consistent way ($5.04\%$/$6.27\%$). Figure \ref{fig:featureImportance} summarizes the importance of the features in this link prediction setting by measuring the \textit{mean decrease Gini}, the average gain of purity achieved when splitting on a given variable (the higher, the better)~\cite{louppe13understanding}. In line with previous work~\cite{yang12predicting}, this analysis further confirms the importance of the temporal features and found that time matters when recommending new contacts.

\vspace{4pt}
\noindent \textbf{Q7: Is it possible to limit the bias of the recommender system while keeping its high accuracy?}

\begin{figure}[t!]
\begin{center}
\includegraphics[width=0.49\columnwidth]{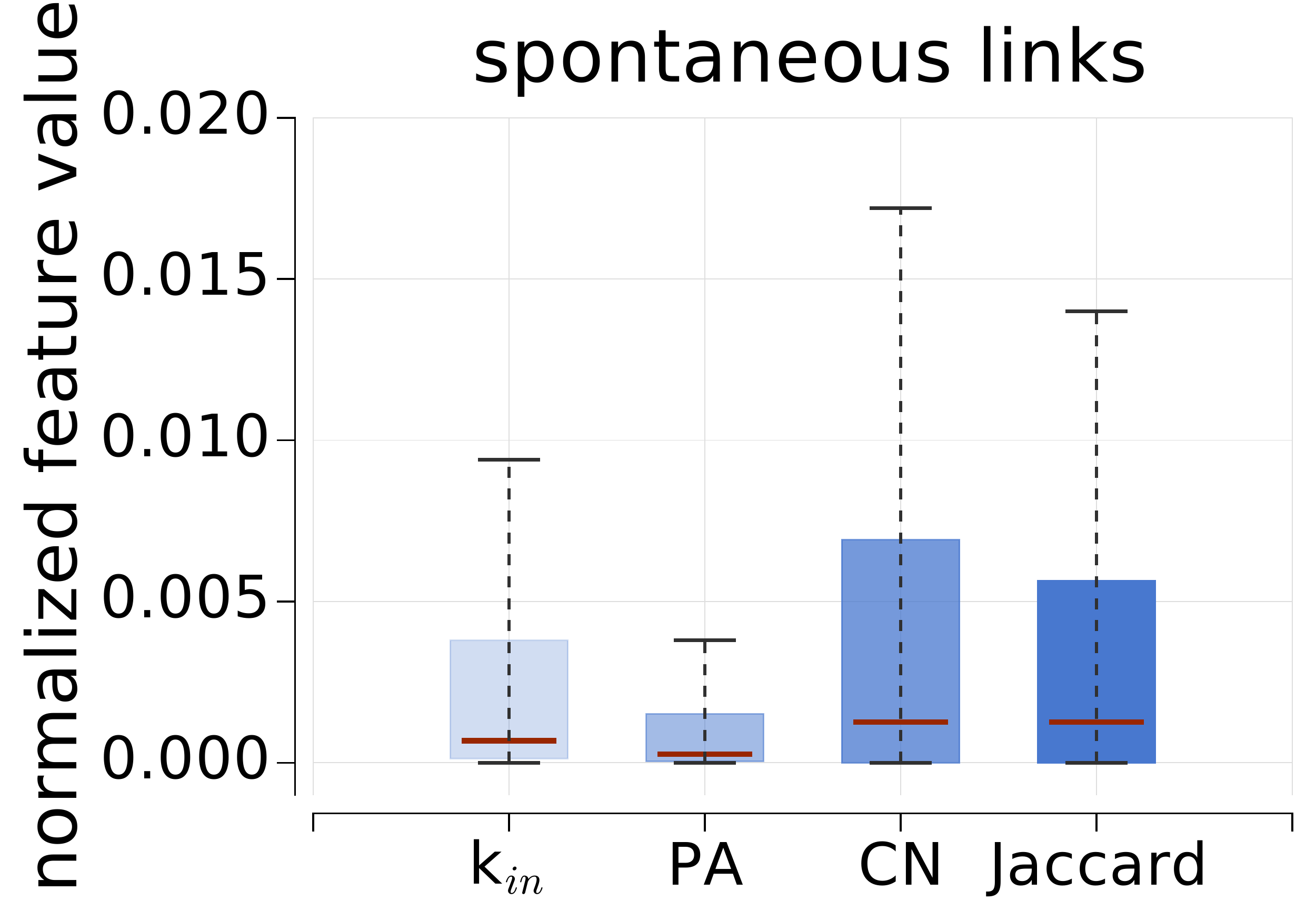}
\includegraphics[width=0.49\columnwidth]{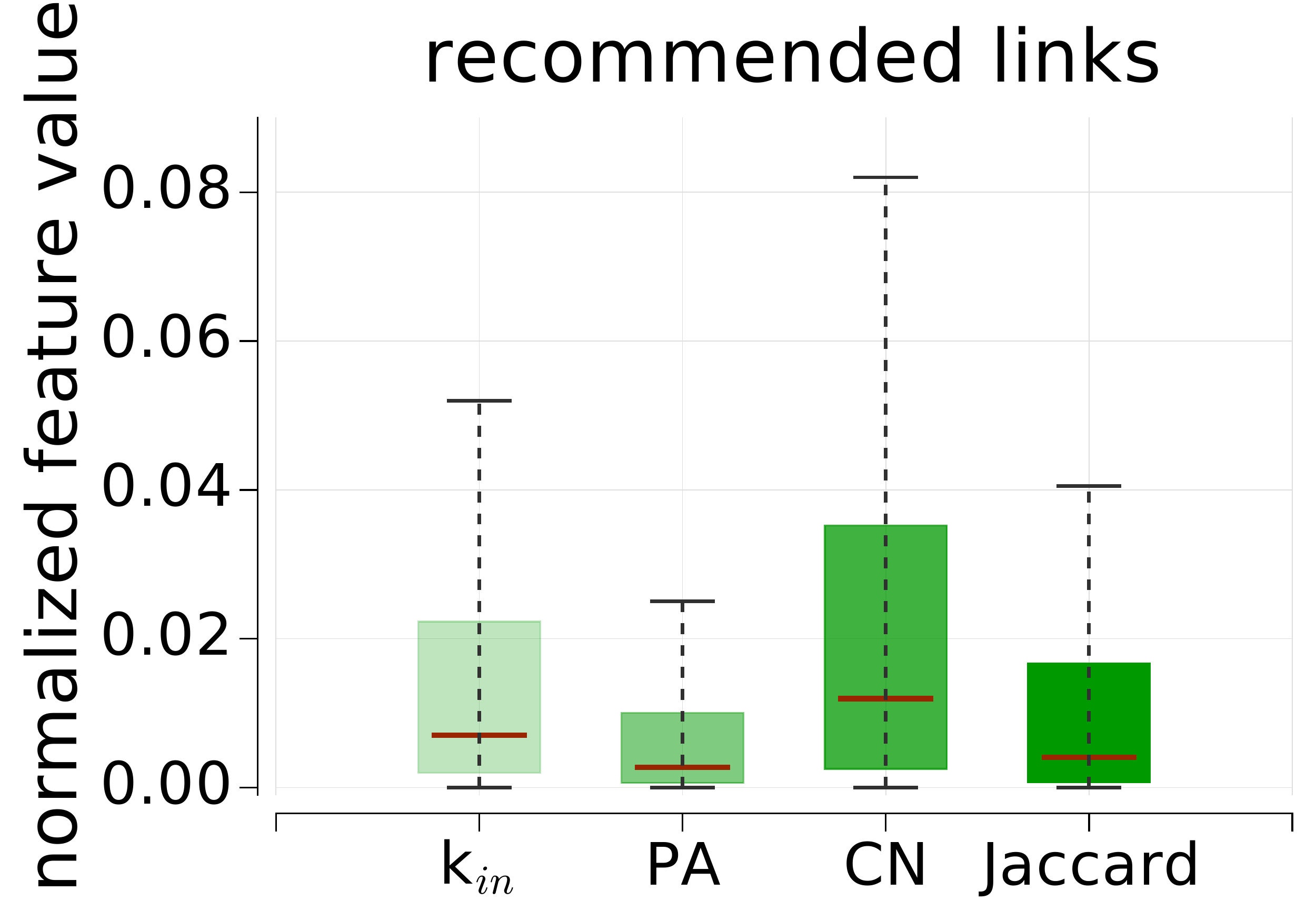}
\caption{Distribution of structural features of spontaneous and recommended links. Given directed links in the form $(i,j)$ we show the boxplots of the distributions of $k_{in}(j)$, $PA(i,j)$, $CN(i,j)$, and Jaccard$(i,j)$.}
\label{fig:feature_analysis}
\vspace{-10pt}
\end{center}
\end{figure}

In Section~\ref{sec:analysis:predictors} we observed that the distribution of nodes' popularity and of their structural similarity with ego are statistically different when considering recommended vs. spontaneous links. To further investigate this difference, we analyze the distribution of all structural features over a sample ($200$k) of recommended and spontaneous links, equally represented. 
For sake of presentation, we normalize the value of each feature on a scale 0 to 1, by dividing the raw feature value by its maximum value in the whole $200$k sample. As shown in Figure~\ref{fig:feature_analysis}, spontaneous links tend to exhibit less degree of structural overlap than recommended links (the median Jaccard value on recommended links is $4$ times larger than the value recorded on spontaneous ones). The same observation holds when analyzing the node popularity; the median in-degree of the target node on recommended links is one order of magnitude higher than the corresponding value computed on spontaneous links. 

Learning to what extent it is possible to automatically tell recommended links and spontaneous links apart would allow us to train new recommender systems to suggest links whose properties better adhere to the natural criteria that people follow when adding new contacts. To gauge this possibility, we run a second prediction experiment with the same features and setup of the previous one but with a different selection of positive and negative examples. 

We pick $100K$ pairs that will be connected in the future through a spontaneous link as positive examples, and as many pairs that will be connected through a recommended link as negative examples. A random forest classifier is able to distinguish the two classes pretty accurately (AUC~$=0.795$, F-measure~$=0.721$).

In a more realistic scenario, the recommender should learn to recognize the space between recommended links and truly negative examples (links that are never formed). To model that, we add to the training set $100K$ negative pairs that will not be connected at any time in the future. This setting achieves a better performance (AUC~$=0.823$, F-measure~$=0.771$) with around $90\%$ accuracy on the negative class and $55\%$ on the positive one. In short, even if very basic structural and temporal features are used, it is possible to effectively use the output of current link recommenders to train new recommenders that smooth the algorithmic bias and produce suggestions that better simulate the spontaneous process of link selection.

\section{Conclusions} \label{sec:conclusions}

We have provided a large-scale analysis of ego-network evolution on two online platforms, exposing the dynamics of their bursty evolution, community-driven growth, diameter expansion, and selection of new nodes based on a time-varying interplay between similarity and popularity. By studying the set of Tumblr links created as a result of algorithmic suggestions, we find that recommended links have different statistical properties than spontaneously-generated ones. We also find evidence that link recommendations foster network diversity by leading nodes that are structurally similar to choose different sets of new neighbors.

Our work has some limitations. Flickr and Tumblr are mainly interest networks, where people follow each other based on topical tastes. Some of the results we report here might not generalize to social networks that aim mostly at connecting people who know each other in real life (e.g., Facebook, LinkedIn). Also, we only consider the network structure and disregard any notion of node profile, including posting activity in time and user-generated content; that information would help to further detail the dynamics of ego-network expansion with respect to other dimensions such as topical similarity between profiles. 

Our results have a number of practical implications. We provide further evidence that in online social networks not all links are created equal; network analysts who produce network growth models based on the observation of online social networks' longitudinal traces should consider weighting links that emerge spontaneously different from those that are created algorithmically. Through our prediction experiments, we also provide a hint about how link recommender systems could incorporate signals on the ego-network's evolutionary stage to improve the quality of suggestions. We hope our work provides yet another step towards a better understanding of the evolutionary dynamics of social networks.

\section{Acknowledgments}
We would like to thank \textbf{Martin Saveski} for his valuable suggestions and the anonymous reviewers for helpful comments.
\balance
\small
\bibliographystyle{abbrv}

\end{document}